\documentclass[runningheads]{llncs}
\usepackage{amsmath}
\usepackage{amssymb}
\usepackage{amsfonts}
\usepackage[T1]{fontenc}
\usepackage{graphicx}
\usepackage{color}

\usepackage{graphicx}%
\usepackage{multirow}%
\usepackage[title]{appendix}%
\usepackage{xcolor}%
\usepackage{textcomp}%
\usepackage{manyfoot}%
\usepackage{booktabs}%
\usepackage{algorithm2e}%
\usepackage{algorithmicx}%
\usepackage{algpseudocode}%
\usepackage{listings}%
\usepackage{tikz}%
\usetikzlibrary{quantikz2}%
\usepackage{svg}
\usepackage{subfig}
\usepackage[hidelinks]{hyperref}
\usepackage[capitalise]{cleveref}
\creflabelformat{equation}{#2#1#3}

\usepackage{todonotes}

\begin{document}
\title{Architectural Influence on Variational Quantum Circuits in Multi-Agent Reinforcement Learning: Evolutionary Strategies for Optimization} %Applying Multi-Agent Quantum Reinforcement Learning using Evolutionary Optimization Strategies
\titlerunning{Architectural Influence on VQC in Evolutionary MARL}
\author{
Michael Kölle\inst{1} \and
Karola Schneider\inst{1} \and 
Sabrina Egger\inst{1} \and 
Felix Topp\inst{1} \and 
Thomy Phan\inst{2} \and
Philipp Altmann\inst{1} \and
Jonas Nüßlein\inst{1} \and 
Claudia Linnhoff-Popien\inst{1}}
\authorrunning{M. Kölle et al.}
\institute{LMU Munich, Oettingenstraße 67, 80538 Munich, Germany \and Thomas Lord Department of Computer Science, University of Southern California, Los Angeles, USA\\
\email{michael.koelle@ifi.lmu.de}}

\maketitle

\begin{abstract}
In recent years, Multi-Agent Reinforcement Learning (MARL) has found application in numerous areas of science and industry, such as autonomous driving, telecommunications, and global health. Nevertheless, MARL suffers from, for instance, an exponential growth of dimensions. Inherent properties of quantum mechanics help to overcome these limitations, e.g., by significantly reducing the number of trainable parameters. 
Previous studies have developed an approach that uses gradient-free quantum Reinforcement Learning and evolutionary optimization for variational quantum circuits (VQCs) to reduce the trainable parameters and avoid barren plateaus as well as vanishing gradients. This leads to a significantly better performance of VQCs compared to classical neural networks with a similar number of trainable parameters and a reduction in the number of parameters by more than 97 \% compared to similarly good neural networks. 
We extend an approach of Kölle et al. by proposing a Gate-Based, a Layer-Based, and a Prototype-Based concept to mutate and recombine VQCs. Our results show the best performance for mutation-only strategies and the Gate-Based approach. In particular, we observe a significantly better score, higher total and own collected coins, as well as a superior own coin rate for the best agent when evaluated in the Coin Game environment.
\end{abstract}

% Contents
\section{Introduction}
\label{sec: introduction}
Artificial intelligence (AI) is currently on everyone's lips. This is due to the fact that it is being used to find innovative solutions in many areas. AI is now a player in industry, healthcare, transportation, and education, for example, as it contributes to the progress of many current technologies \cite{jiang2022quo}. 

But not only single-agent settings are needed; Multi-Agent Systems (MAS) are also central elements in these application areas. Although these agents are inherently designed to act independently, they can be orchestrated to cooperate effectively through Multi-Agent Reinforcement Learning (MARL). MARL has shown significant efficacy, particularly in addressing social dilemmas \cite{leibo2017multi}. 

Reinforcement Learning (RL) algorithms, which constitute a subdomain of AI, are already able to outperform humans in domains like video games \cite{badia2020agent57}, \cite{schrittwieser2020mastering}. At the same time, quantum computing is an emerging technology that allows for faster problem-solving and more efficient training in RL \cite{harrow2017quantum}. Nevertheless, Quantum Reinforcement Learning faces new challenges, for example barren plateaus or vanishing gradients \cite{franz2023uncovering}, \cite{chen2022variational}. Research faces these problems, e.g., by using evolutionary optimization methods that provide promising results \cite{chen2022variational}. This suggested the combination of MARL and quantum techniques, which leads to Multi-Agent Quantum Reinforcement Learning (MAQRL).

In this study, we employ Variational Quantum Circuits (VQCs) to model each agent, leveraging their versatility and capability to encode complex information. VQCs are particularly suitable for representing agent behaviors in varied environments due to their inherent flexibility. To optimize the performance of these quantum circuits, we utilize an evolutionary algorithm, which iteratively adjusts the VQC parameters to enhance decision-making and interactions. While evolutionary algorithms are traditionally used in classical contexts, applying them to the quantum domain opens up new possibilities for efficiently exploring the extensive parameter space of VQCs.

To find out if and how evolutionary optimization can be used in MAQRL, we evaluate multiple generational evolutionary strategies. We compare the results with each other to find out which approaches work best to leverage superior performance from the VQCs. For robustness and fairness, all experiments are conducted in a cooperative Coin Game setting, as it is a regarded Multi-Agent environment.

Our contribution includes the application of evolutionary optimization in a quantum Multi-Agent Reinforcement Learning setting. We introduce three evolutionary strategies for Variational Quantum Circuits and evaluate them in a Reinforcement Learning setting. Furthermore, we compare the results gained from these experiments with a static baseline. 

This work is structured as follows: in \cref{sec: preliminaries}, we give a short introduction to Multi-Agent Reinforcement Learning, Evolutionary Optimization, Quantum Computing, and the structure of Variational Quantum Circuits. Afterwards, we highlight some work that is related to our study (\cref{sec: related}). Then, we outline the approach by K{\"o}lle et al. \cite{kolle2023multi} before we introduce our approach (\cref{sec: approach}). In \cref{sec: experimental}, we provide information about our experimental setup, and in \cref{sec: results}, we share the results of our study. We conclude in \cref{sec: conclusion} with a short summary and suggestions for further research. All experiment data and code can be found here \footnote{https://github.com/karolaschneider/vqc-opt-arch-evo}.
\section{Preliminaries}
\label{sec: preliminaries}
In this section, we provide foundational concepts necessary for understanding the multi-agent setting and the methodologies employed in our research. We begin by discussing the framework of Markov games, which underpins the interactions among multiple agents. This is followed by an overview of independent learning in MARL, highlighting the challenges and dynamics involved. Next, we delve into the specifics of evolutionary optimization and its application in MARL, drawing inspiration from natural selection to improve agent performance. Finally, we cover the basics of quantum computing and VQCs, which are integral to our approach.

\subsection{Multi-Agent Setting}
We focus on \textit{Markov games} $M=\langle D,S,A,P,R \rangle$, where $D=\{1,\ldots,N\}$ represents a set of agents $i$, $S$ is a set of states $s_t$ at time step $t$, and $A=\langle A_1,\ldots,A_N \rangle$ is the set of joint actions $a_t= \langle a_{t,i} \rangle_{i \in D}$. The transition probability is denoted as $P(s_{t+1}\vert s_t,a_t)$, and the joint reward is $\langle r_{t,1}, \ldots, r_{t,N} \rangle =R(s_t,a_t) \in \mathbb{R}$. The action selection probability for agent $i$ is represented by the individual policy $\pi(a_{t,i} \vert s_t)$.

The policy $\pi_i$ is typically evaluated using a \textit{value function} $V^\pi_i(s_t)=\mathbb{E}_\pi[G_{t,i}\vert s_t]$ for all $s_t \in S$, where $G_{t,i}= \sum_{k=0}^\infty \gamma^k r_{t+k,i}$ represents the individual and discounted \textit{return} of agent $i \in D$ with a discount factor $\gamma \in [0,1)$, and $\pi=\langle\pi_1, \ldots,\pi_N\rangle$ is the \textit{joint policy} of the multi-agent system (MAS). The goal for agent $i$ is to find the \textit{best response} $\pi_i^*$ with $V_i^* = max_{\pi_i}V_i^{\langle\pi_i,\pi_{-i}\rangle}$ for all $s_t \in S$, where $\pi_{-i}$ denotes the joint policy \textit{excluding} agent $i$.

We define the \textit{efficiency} of a MAS or \textit{utilitarian metric} $(U)$ by the sum of all individual rewards until time step $T$:
\begin{equation} \label{eq: utilitarian}
    U = \sum \limits_{i\in D}R_i,
\end{equation}

where $R_i=\sum_{t=0}^{T-1}r_{t,i}$ is the \textit{undiscounted return} or the \textit{sum of rewards} for agent $i$ starting from the initial state $s_0$.

\subsection{Multi-Agent Reinforcement Learning}
In our research, we focus on independent learning, where each agent $i$ optimizes its individual policy $\pi_i$ based on its own information, such as $a_{i,j}$ and $r_{i,j}$, using reinforcement learning (RL) techniques. For instance, we employ evolutionary optimization as described in \cref{sec: evo_opt} on Evolutionary Optimization.

Independent learning introduces non-stationarity because the agents adapt simultaneously, continuously altering the environment dynamics from each agent's perspective \cite{littman1994markov}, \cite{laurent2011world}, \cite{hernandez2017survey}. This non-stationarity can lead to the development of overly greedy and exploitative policies, causing agents to defect from cooperative behavior \cite{leibo2017multi}, \cite{foerster2017learning}.

\subsection{Evolutionary Optimization} \label{sec: evo_opt}
Inspired by the process of natural selection, evolutionary optimization has demonstrated efficacy in solving complex problems where traditional methods may fall short \cite{vikhar2016evolutionary}. This approach utilizes a population of individuals, each randomly generated with a unique set of parameters. These individuals are assessed based on a fitness function that evaluates their performance on the given problem. The most fit individuals are then selected for reproduction, with their parameters recombined and mutated to create a new population for the next generation \cite{eiben2015introduction}.

Evolutionary optimization techniques, such as genetic algorithms \cite{holland1991artificial}, have been successfully applied in various fields. These include the optimization of neural networks and interactive recommendation tasks \cite{ding2011optimizing}, \cite{gabor2019benchmarking}. Moreover, these methods have been employed to address a broad spectrum of challenges, from designing quantum circuit architectures to optimizing complex real-world designs \cite{lukac2002evolving}, \cite{caldas2002design}.

\subsection{Quantum Computing}
Quantum computing is an emerging field in computer science that leverages the principles of quantum mechanics to process information. Unlike classical computers, which use bits to store and process data, quantum computers use quantum bits, or qubits, which can exist in multiple states simultaneously due to the principle of \textit{superposition} \cite{yanofsky2008quantum}.

A qubit's state, denoted as $\ket{\Psi}$, can be expressed as a linear combination of the basis states $\ket{0}$ and $\ket{1}$:
\begin{equation}
    \ket{\Psi} = \alpha \ket{0} + \beta \ket{1},
\end{equation}

where $\alpha$ and $\beta$ are complex coefficients that satisfy the normalization condition:
\begin{equation}
    \lvert \alpha \rvert^2 + \lvert \beta \rvert^2 = 1.
\end{equation}

Upon measurement, a qubit in the superposition state $\alpha \ket{0} + \beta \ket{1}$ collapses to one of the basis states, $\ket{0}$ or $\ket{1}$, with probabilities determined by the magnitudes of the coefficients, $\lvert \alpha \rvert^2$ and $\lvert \beta \rvert^2$, respectively. This process marks the transition from a quantum superposition to a definite classical state, where the observable's value is precisely known \cite{nielsen2010quantum}.

Moreover, multiple qubits can be entangled, creating strong correlations between them. \textit{Entanglement} allows the qubits to be interconnected in such a way that the state of one qubit directly influences the state of another, regardless of the distance separating them. This property is pivotal for the enhanced computational power of quantum systems, enabling them to solve complex problems more efficiently than classical computers.

\subsection{Variational Quantum Circuits}
Variational Quantum Circuits (VQCs), also known as parameterized quantum circuits, are quantum algorithms designed as function approximators that are trained using classical optimization methods. They are increasingly used as alternatives to neural networks in Deep Reinforcement Learning (Deep RL) applications \cite{chen2022variational}, \cite{schuld2020circuit}, \cite{chen2020variationalquantumcircuitsdeep},\cite{skolik2021layerwise}, \cite{chen2023quantum}. A VQC typically comprises three main stages: State Preparation, Variational Layers, and Measurement.

In the state preparation phase, classical input data is encoded into a quantum state through superposition. We employ Amplitude Embedding \cite{mottonen2004transformation} to embed classical data into the amplitudes of qubits. For instance, to embed a feature vector $x \in \mathbb{R}^3$ into a 2-qubit quantum state $\Psi = \alpha \ket{00} + \beta \ket{01} + \gamma \ket{10} + \delta \ket{11}$, where  $\lvert \alpha \rvert^2 + \lvert \beta \rvert^2 + \lvert \gamma \rvert^2 + \lvert \delta \rvert^2 = 1$, the feature vector is first padded to match $2^n$ features (with $n$ being the number of qubits). Next, the padded feature vector $y$ is normalized such that
$\sum_{k=0}^{2^n-1} \frac{y_k}{\lvert\lvert y \rvert \rvert} = 1$. Finally, we use the state preparation method by Mottonen et al. \cite{mottonen2004transformation} to encode the normalized feature vector into the qubit state.

\begin{figure}
    \centering
%    \begin{adjustbox}{width=\linewidth}
    \begin{quantikz}
    \lstick[wires=6]{$\ket{0}$} & \gate[wires=6]{U(x)} & \ctrl{1}\gategroup[6,steps= 8,style={dashed,
    rounded corners,fill=blue!20, inner xsep=2pt},
    background]{{\sc Variational Layer}} & \qw & \qw & \qw & \qw &  \targ{} & \qw & \gate{R(\alpha _1, \beta _1, \gamma _1)}& \meter{} && \\
    & \qw &\targ{} & \ctrl{1} & \qw &  \qw & \qw & \qw & \qw &\gate{R(\alpha _2, \beta _2, \gamma _2)} & \meter{} &&\\
    & \qw & \qw & \targ{} & \ctrl{1} & \qw & \qw & \qw & \qw & \gate{R(\alpha _3, \beta _3, \gamma _3)} & \meter{} && \\
    & \qw & \qw & \qw & \targ{} & \ctrl{1} \qw & \qw & \qw & \qw &\gate{R(\alpha _4, \beta _4, \gamma _4)}&\meter{} &&\\
    & \qw & \qw & \qw & \qw & \targ{} & \ctrl{1} & \qw & \qw & \gate{R(\alpha _5, \beta _5, \gamma _5)} \\
    & \qw & \qw & \qw & \qw & \qw & \targ{} & \ctrl{-5}& \qw & \gate{R(\alpha _6, \beta _6, \gamma _6)}
    \end{quantikz}
%    \end{adjustbox}
    \caption{Variational Quantum Circuit used by K{\"o}lle et al. \cite{kolle2023multi}}
    \label{fig:usedvqc}
\end{figure}
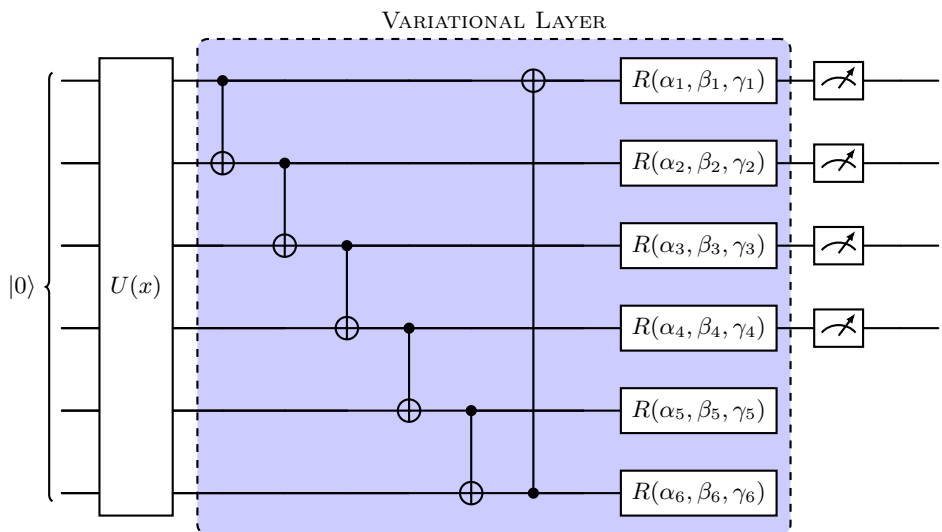

The variational layers stage involves repeated application of single-qubit rotations and entangling gates. Drawing inspiration from the circuit-centric classifier design \cite{schuld2020circuit}, our circuits incorporate three single-qubit rotation gates (denoted as $\theta_i^j$, where $i$ is the qubit index and $j \in \{0,1,2\}$ is the index of the single-qubit rotation gate) and CNOT gates as entanglers. The architecture within the dashed blue area in \cref{fig:usedvqc} is repeated $L$ times, where $L$ represents the number of layers. The target qubit for the CNOT gate in each layer is determined by $(i+l) mod$ $n$, with $l$ indicating the current layer.

In the final stage, the expectation value is measured in the computational basis ($z$) for the first $k$ qubits, where $k$ corresponds to the dimensions of the agents’ action space. Each measured expectation value is assigned a bias, which is included in the VQC parameters and updated during training.

These stages enable VQCs to effectively approximate complex functions, making them suitable for various quantum machine learning and optimization tasks. The adaptability of VQCs in modifying their parameters and structures allows them to tackle computationally demanding problems that classical algorithms struggle to solve.
\section{Related Work}
\label{sec: related}
Our work extends the approach for using evolutionary optimization for quantum Reinforcement Learning using VQCs by K{\"o}lle et al. \cite{kolle2023multi}. They, in turn, were inspired by Chen et al. \cite{chen2022variational} who showed that it is possible to reduce parameter requirement in comparison to classical RL when leveraging VQCs for Q-value function approximation in Deep Q-Learning. 
K{\"o}lle et al. \cite{kolle2023multi} then developed three gradient-free genetic variations of VQCs with MAQRL applying evolutionary optimization. They compared the results of their agents with Neural Networks in the Coin Game \cite{lerer2017maintaining}, finding similar accomplishments with more than 97 \% less parameters within the VQC.

Another approach to Quantum Multi-Agent Reinforcement Learning is leveraging Quantum Boltzmann Machines (QBM) \cite{neumann2020multi}, \cite{muller2021towards}. This strategy builds on an existing methodology where QBM outperforms classical Deep RL in terms of convergence speed, measured by the number of required time steps while using Q-value approximation. The results suggest enhanced stability in learning and the agents' ability to attain optimal strategies in grid domains. Similarities to the method employed here lie in the utilization of Q-values and grid domains as testing environments.

Yun et al. \cite{yun2022quantum} employ Quantum Multi-Agent Reinforcement Learning using centralized learning and decentralized execution to avoid the challenges of the NISQ era and the non-stationary properties from classic MARL. This approach achieves a superior overall reward in the tested environments compared to classical methodologies using less parameters. 

\section{Approach}
\label{sec: approach}

In this section we describe an idea of optimizing a $\theta$ parameterized agent using the evolutionary approach by K{\"o}lle et al. \cite{kolle2023multi}. In the next step, we extend this concept by applying architectural changes and parameter adjustments to explore the impact of circuit architecture on the performance.

K{\"o}lle et al. \cite{kolle2023multi} were inspired by Chen et al. \cite{chen2022variational}, but in contrast, they use a more general Multi-Agent setup to optimize the utilitarian metric $U$ (\cref{eq: utilitarian}). For maximizing the fitness function, they use a population $P$, which consists of $\eta$ randomly initialized agents. To fully match the quantum circuits' parameter space, the agents are parameterized by $\theta \in [-\pi,\pi)$. 

Contrary to past work, they use a Variational Quantum Circuit (VQC) instead of neural networks for approximating the value of the agent's action. This primarily aims to demonstrate the improved parameter efficiency, even when applied to complex learning tasks. A VQC consists mainly of three components: the input embedding, the repeated variational layers, and the measurement. \cref{fig:usedvqc} illustrates the VQC employed by \cite{kolle2023multi}.

First, it is necessary to translate the classical data into a quantum state which can be achieved by using different embedding strategies like Basis Embedding, Angle Embedding or Amplitude Embedding. At the moment, Amplitude Embeddings is, considering the high dimensionality of the state space, the only viable embedding strategy that can embed the whole state information, enabling the embedding of $2^{n_q}$ states in $n_q$ qubits.

Variational layers constitute the second part of VQCs and are variably repeated. Each iteration increases the number of $\alpha_i, \beta_i, \gamma_i$ parameters, which are defined by $\theta$ and depict each individual that should be optimized. In consequence, each layer has $n_\theta=n_q *3$ parameters, where $n_q$ is the number of qubits. All rotations are performed sequentially as $R_Z(\alpha_i)$, $R_Y(\beta_i)$, and $R_Z(\gamma_i)$. Before the parameterized rotations can be applied, each variational layer includes adjacent CNOT gates to entangle all qubits. 

After $n_l$ repetitions of the variational layer, the third part follows: we can measure the first $n_a$ qubits, where $n_a$ is the number of possible actions, to determine the predicted values of the individual actions. A Z-axis measurement is used to determine the Q-value of the corresponding action, and an agent chooses the action with the greatest expected value.

To optimize the utilitarian metric $U$ (\cref{eq: utilitarian}), K{\"o}lle et al. \cite{kolle2023multi} train the parameters $\theta$ with the help of an evolutionary algorithm which works as follows: First, compute the fitness of each individual $i$ by performing $\kappa$ steps in the environment for each generation. Based on this fitness, select the $\tau$ best agents for the development of the upcoming generation.

Second, for building a new population in the next generation, join mutation and recombination possibilities. Through crossover, recombine the $\tau$ best agents of the current generation to new individuals. The descendants are the result of a random selection of two parents and then crossing their parameters at a randomly selected index. Alternatively, with a mutation step, new agents are gained. Here, the parameters $\theta$ of the $\tau$ best agents in the current generation are modified as follows:
\begin{equation} \label{eq: modify}
    \theta = \theta + \sigma * \varepsilon
\end{equation}

Again, the agents with the best fitness value $T$ of the new generation will be the parents of the next upcoming generation. For the mutation, the parameters $\theta$ are modified like in \cref{eq: modify} with the mutation power $\sigma$ and the Gaussian noise $\epsilon \sim \mathcal{N(0,1)}$. As a consequence, all $\theta_i$ parameters undergo minor mutations, and new agents, called children, are generated.

\begin{figure}[htb]
    \centering
    \includegraphics[width=\linewidth]{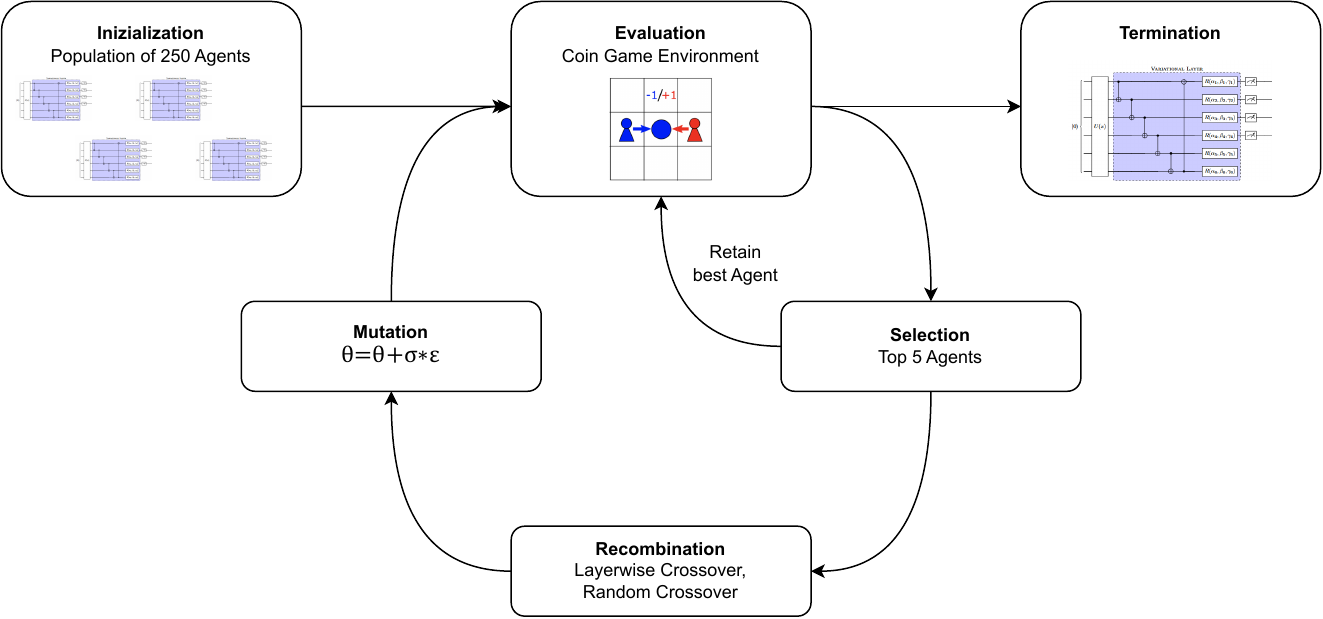}
    \caption{Training Loop of Kölle \textit{et al.} \cite{kolle2023multi}}
    \label{fig: process}
\end{figure}

The third and last step in every generation is to add the unaltered elite agent to the child population.
\cref{alg: evo} shows the EA training procedure proposed by K{\"o}lle et. al. \cite{kolle2023multi} for finding the utilitarian metric $U$ (\cref{eq: utilitarian}) and \cref{fig: process} gives a graphical overview of the training procedure. 

\SetKwComment{Comment}{/* }{ */}

\begin{algorithm}[hbt] 
\caption{Evolutionary Optimization Algorithm by K{\"o}lle et al. \cite{kolle2023multi}}
\label{alg: evo}
\KwData{Number of Generations $\mu$, Population Size $\eta$, Evaluation Steps $\kappa$, Truncation Selection $\tau$, Mutation Power $\sigma$, and Number of Agents $N$}
\KwResult{Population $\mathcal{P}$ of optimized agents}
$\mathcal{P}_0 \gets$ Initialize population $\eta$ with random $\theta$\;
\For{$g \in \{0, 1, ..., \mu-1\}$}{
    \For{$ i \in \{0, 1, ..., \eta-1\}$}{
    Reset testing environment \\
    Score $S_{t=0, i} \gets 0$ \\
    \For{$t \in \{0, 1, ..., \kappa-1\}$}{
    Use policy of agent $i$ for all agents in env \\
    Select action $a_t \gets argmax VQC_\theta(s_t)$ \\
    Execute environment step with action $a_t$ \\
    Observe reward $r_t$ and next state $s_{t+1}$ \\
    $S_{t,i} \gets S_{t-1,i} + r_t$ \\
    }
    }
    $\lambda \gets$ Select top $\tau$ agents based on $S_{\kappa-1,i}$ \\
    Keep top agent based on $S_{\kappa-1,i}$ \\
    Recombine $\eta - 1$ new agents out of $\lambda$ \\
    Mutate $\eta - 1$ generated agents \\
    $\mathcal{P}_{g+1} \gets \eta - 1$ generated agents + top agent \\
}
\end{algorithm}

Now we extend this approach by using different architectural concepts for the variational layers and additionally, apply different techniques for the evolutionary alternation of the circuit's architecture: a Layer-Based concept, a Gate-Based concept and a hybrid Prototype-Based concept.

\subsection{Layer-Based Concept}

The Layer-Based concept is inspired by Strongly Entangling Layers \cite{schuld2020circuit,giovagnoli2023qneat}, where all qubits are entangled. This is realized by applying a CNOT Gate to every qubit $q_n$ and selecting qubit $q_{n-1} mod $ $n$ as the target qubit. Then $R_X$, $R_Y$ and $R_Z$ rotation gates are performed for every qubit. In contrast to \cite{kolle2023multi}, during the evolutionary optimization process the number of layers can increase and decrease due to mutation and recombination.
Here, the evolutionary algorithm creates new circuits by recombining the initial layers of one circuit with the final layers of another. Mutation alters the circuit's design by either adding or removing entire layers, thereby preserving the overall layer structure.
During tests, we found that the evolutionary optimization process works best when starting with circuits that consist of only one layer.

\subsection{Gate-Based Concept}

When applying the Gate-Based concept, we eliminate the organization of gates into layers. Instead, the circuit is built by applying a specific number of randomly sampled gates to randomly chosen qubits. These gates are selected from a predefined gate set ($R_X$, $R_Y$, $R_Z$, and CNOT), and are placed on a qubit $q$ within the range $[0, n_q - 1)$, where $n_q$ represents the number of qubits in the VQC. For multi-qubit gates like the CNOT gate, an additional distinct qubit is selected to serve as the control qubit.
During the evolutionary step, adjustments are made at the gate level. This includes deleting, adding, or substituting gates, as well as creating new circuits by recombining gates from two parent circuits.

\subsection{Prototype-Based Concept}

The Prototype-Based concept merges the Layer-Based and Gate-Based approaches by creating a circuit with repeated layers, while manipulations occur at the gate level. One layer of this circuit is constructed similarly to a Gate-Based Circuit, and the resulting gate arrangement, called the prototype, is then repeated across all layers of the circuit. The evolutionary processes applied to the prototype work the same as for the Gate-Based approach. This approach ensures, that the gate composition remains consistent for all layers.

As we change the circuit architecture, we need to adjust some parameters for the initialization of VQCs: Layer-Based circuits have a specific number of initial layers $\upsilon$, Gate-Based circuits have a specific number of initial gates $\chi$, and Prototype-Based circuits have a specific number of initial layers $\upsilon$ as well as a specific number of initial gates per layer $\chi$. This means when constructing a Gate-Based or a Prototype-Based circuit, we have to randomly select $\chi$ gates from our predefined gate set ($R_X, R_Y, R_Z, CNOT$) and place them on $\chi$ randomly selected qubits within the circuit. For Prototype-Based circuits, this procedure is repeated $\upsilon$ times. When construction a Layer-Based circuit, we need $\upsilon$ layers in the predefined structure and randomly initialize the parameters of the gates within the range $[-\pi, \pi)$ to cover the whole parameter space of the circuit.   

\subsection{Evolutionary Algorithm}

\begin{figure}[htb]
    \centering
    \includegraphics[width=\linewidth]{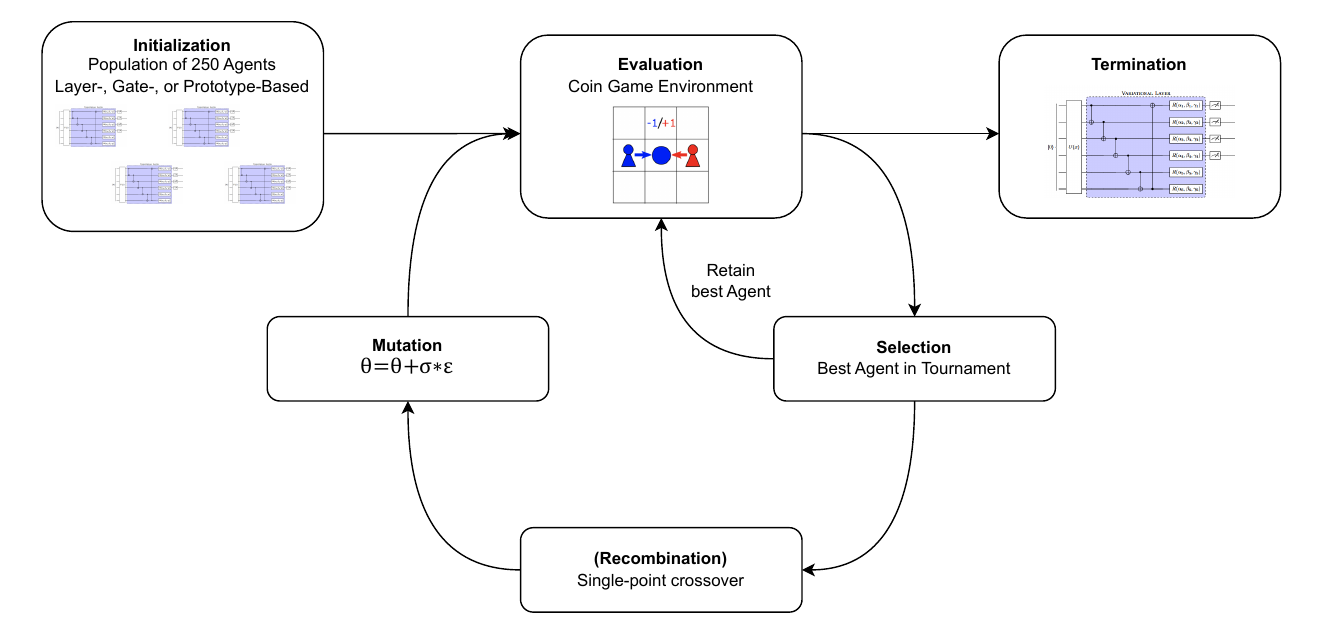}
    \caption{Training loop of approach containing architectural changes}
    \label{fig: process-arch}
\end{figure}

\begin{algorithm}[hbt]
\caption{Evolutionary Optimization Algorithm of approach containing architectural changes}
\label{alg: evo-arch}
\KwData{Number of Generations \(\mu\),
Population Size \(\eta\),
Evaluation Steps \(\kappa\),
Tournament Size \(\tau\),
Mutation Rate \(\phi\),
Parameters \(\theta\),
and Number of Agents in game \(N\)}
\KwResult{Population \(P\) containing an optimized agent}
$\mathcal{P}_0 \gets$ Initialize population $\eta$ with random $\theta$\;
\For{$g \in \{0, 1, ..., \mu-1\}$}{
    \For{$ i \in \{0, 1, ..., \eta-1\}$}{
        Reset testing environment \\
        Score $S_{t=0,i} \gets \text{0}$ \\
        \For{$t \in \{1, 2, \ldots, \kappa\}$} {
            $j \gets \text{iterate through }\{1, 2, \ldots, N\}$ \\
            $\text{Select action } a_t \gets argmax \, VQC_{\theta i}(s_t)$ \\
            $\text{Execute action } a_t \text{ for agent } j$ \\
            $\text{Observe reward } r_t \text{ and next state } s_{t+1}$ \\
            $S_{t,i} \gets S_{t-1,i}+r_t$ \\
        }
    }
    $\alpha \gets \text{best agent based on } S_{\kappa-1 ,i} \text{ and circuit size}$ \\
    $\text{Recombine }\eta -1 \text{ new agents using tournament selection with size }\tau$ \\
    $\text{Mutate }\eta -1 \text{ generated agents according to mutation rate }\phi$ \\
    $P_{g+1} \gets \eta -1 \text{ generated agents }+\alpha$ \\
}
\end{algorithm}

Like in K{\"o}lle et al. \cite{kolle2023multi}, for all of the three aforementioned concepts, two different strategies for creating the next generation can be used: (1) recombination and mutation, where new individuals are formed by recombining selected agents from the existing population. Following this, they are mutated based on the mutation rate. (2) mutation only, where new individuals are created by mutating selected agents from the current population. Both approaches lead to a new generation of population size $\eta$, including the elite agent from the previous generation. \cref{alg: evo-arch} shows the employed EA, that extends the approach by K{\"o}lle et al. \cite{kolle2023multi} by applying architectural changes and \cref{fig: process-arch} depicts the training process. 
In detail, this works as follows: 

We use tournament \textbf{selection} to choose individuals from the population. This process starts by randomly picking $\tau$ individuals from the population to form a tournament. The fitness of each individual in the tournament is then compared, and the one with the highest fitness is selected \cite{sunkel2023ga4qco}. By increasing the size of the tournament, the likelihood of selecting high-performing agents rises, thereby boosting the selection pressure \cite{eiben2015introduction}.

We implement \textbf{elitism} by carrying the unchanged fittest agent into the next generation \cite{chen2022variational}. This strategy ensures that advantageous traits are preserved, preventing a decline in performance and promoting continuous improvement within the population \cite{grocsan2003role}.

\textbf{Recombination} generates new individuals, also called offspring, by merging the genetic information of existing individuals, called parents \cite{eiben2015introduction}. In our study, this genetic information includes both the architecture and parameters of the VQCs. We utilize a variant of single-point crossover for this purpose, where the genetic data from two parent individuals is split at a specific point and then rearranged to form new offspring \cite{sunkel2023ga4qco}. This process aims to carry forward beneficial traits from the parents while exploring the solution space \cite{eiben2015introduction}. We choose a cut point that is valid and positioned between layers (Layer-Based concept) or gates (Gate-Based and Prototype-Based concepts). This cut point must fall within the range of both parent circuits and allow each circuit to be divided into a front and a back section, preventing splits at the very beginning or end. In cases where a circuit contains only a single layer or gate, the cut point is positioned just after that layer or gate. This precaution ensures that recombination does not yield an empty circuit, even when merging two single-layer or single-gate circuits. Two offspring are produced by merging the initial segment of one parent circuit with the final segment of the other parent circuit. For Prototype-Based circuits, the recombination is applied to the prototype analogously to Gate-Based circuits and offspring is created by repeating the recombination $n$ times to retain the layer concept. Furthermore, the parameters of each gate are also recombined in the same way.

The \textbf{mutation} modifies an individual's genetic information \cite{eiben2015introduction}. The degree of these alterations can be regulated using the mutation power $\sigma$ \cite{chen2022variational}, while the mutation rate $\phi$ controls how frequently mutations are applied \cite{eiben2015introduction}. In our study, both the parameters and the architecture of a VQC are subject to mutation. The process of parameter mutation is consistent across all architectural variations, including the approach by K{\"o}le et al. \cite{kolle2023multi}. The process of architectural mutation with mutation rate $\sigma_a$ differs depending on the circuits' concept: in Layer-Based circuits mutations add or delete whole layers, where the addition can happen at any position before and after an existing layer. However, it is necessary to meet the rules for layer composition. In Gate-Based circuits mutations are able to remove or replace an existing gate, or to insert a new gate at any position within the circuit. Again, this is restricted to following the rules for placing gates in the circuit. In Prototype-Based circuits, mutations occur in the same manner as in Gate-Based circuits, affecting the prototype of the circuit according to the same rules.
\section{Experimental Setup}
\label{sec: experimental}
In this section, we describe details about our experimental setup we use to evaluate the evolutionary algorithm introduced in \cref{sec: approach}. We apply the aforementioned strategies for creating different architectural circuits. The following parts include information about our game environment, baselines, metrics, and hyperparameters we use for training.

\subsection{Coin Game Environment}
We use the Coin Game \cite{lerer2017maintaining} environment in a 3x3 gridworld version as it has a comparatively small observation space. The use of the short evaluation  environment is due to the fact that we are currently in the Noisy Intermediate-Scale Quantum (NISQ) era, where we can only simulate a restricted amount of qubits, thus limiting the data we can embed into a quantum circuit \cite{preskill2018quantum}.

\begin{figure}
    \centering
    \includegraphics[width=0.4\linewidth]{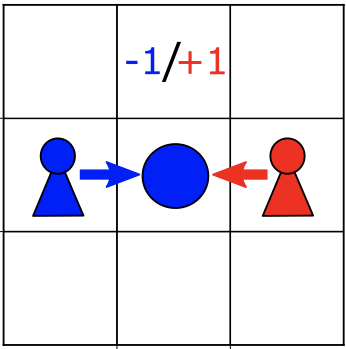}
    \caption{Example State of the Coin Game \cite{phan2022emergent}.}
    \label{fig: CoinGame}
\end{figure}

In the Coin Game, both the Red and Blue agents aim for collecting coins, with a single coin placed on a free cell in the grid. The color of the coin always corresponds to one of the agents' colors. An example state of this well-known sequential game that is suited for evaluating RL strategies is illustrated in \cref{fig: CoinGame}.

When an agent occupies the same position as a coin, the coin is collected. Once a coin is collected, a new one appears at a random location that is not occupied by an agent, and it is again either red or blue. Each game of the Coin Game lasts for 50 steps, with each agent getting 25 turns. The goal is to maximize the agents' rewards.

The Coin Game can be played in both competitive and cooperative modes. In the cooperative mode, the agent that collects a coin receives an reward of +1. Conversely, the second agent's reward decreases by -2 if the first agent collects a coin of their color. Therefore, considering the total reward, collecting one's own coin results in a common reward of +1, while collecting an opponent's coin leads to -1. This setup encourages agents to collect their own coins and leave the opponent's coins for the other agent to collect. If both agents act randomly, the expected reward is zero, making the Coin Game a zero-sum game.

Each cell in the 3×3 gridworld can contain agent 1, agent 2, a red coin, or a blue coin. Empty cells are not included in the observation since agents can move into them freely. Agents can choose from four possible actions, provided the move does not take them outside the 3×3 grid. The actions are numerically coded: 0 for a step north, 1 for a step south, 2 for a step west, and 3 for a step east. An action is considered legal and thus can be executed by a player if the player does not exit the board and the field is unoccupied or contains a coin. To ensure that the VQC does not select illegal actions, the expected values are normalized to the range $[0,1]$ and are masked according to the rules of the environment.

\subsection{Baselines}

VQCs serve as an alternative to classical neural networks in agent design. In the first part of the evaluation, neural networks are used as agents because they function as general approximators. The basic neural network consists of 2-layers for this purpose. The first layer maps the input observations to a variable number of hidden units $x$, and the second layer connects these hidden units to the number of possible actions. This configuration yields the individual Q-values for each action, similar to the VQC approach. To prevent the selection of illegal actions, the Q-values are adjusted using an action mask. Our focus is on this specific neural network with varying hidden unit counts, so that the network can be modified in many ways to affect the outcomes.

In the second part of our evaluation we use a static baseline, which consists of a 8 layers of the Layer-Based approach where we only change the parameters and not the architecture. For the baseline, we run a mutation-only strategy with parameter mutation power $\sigma_p = 0.01$.

\subsection{Metrics}
We evaluate our experiments in the Coin Game environment using three metrics: Score, Total Collected Coins, and Own Coin Rate. In our study, the agents play solely against themselves to simplify the evaluation of their performance.
\begin{enumerate}
    \item \textbf{Score ($S_n$)}: This metric consists of the undiscounted individual rewards $r_{t,i}$ accumulated over all agents until timestep $T \in \{0..49\}$. It is calculated as:
    \begin{equation}
        S_n = \sum\limits_{i \in \{0, 1\}}^{} \sum\limits_{t=0}^{T-1} r_{t,i}
    \end{equation}
    Here, $i$ represents the agent and $n$ (ranging from 0 to 199) denotes the generation, averaged over five seeds. This score provides a comprehensive indicator of the agents' performance in the Coin Game environment.
    \item \textbf{Total Coins Collected ($TC_n$)}: This metric tracks the total number of coins collected $c_{t,i}$ by all agents until timestep $T \in \{0..49\}$:
    \begin{equation}
        TC_n = \sum\limits_{i \in \{0,1\}} \sum\limits_{t=0}^{T-1}c_{t,i}
    \end{equation}
    Similar to the Score metric, $i$ represents the agent and $n \in \{0..199\}$ the generation, averaged over five seeds. This metric helps in understanding the total collection performance of the agents.
    \item \textbf{Own Coins Collected ($OC_n$)}: This metric is the sum of all collected coins that match the agent's own color $o_{t,i}$ until timestep $T \in \{0..49\}$ accumulated over all agents:
    \begin{equation}
        OC_n = \sum\limits_{i \in \{0,1\}} \sum\limits_{t=0}^{T-1}o_{t,i}
    \end{equation}
    Again, similar to the previous metrics, $i$ represents the agents and $n \in \{0..199\}$ the generation, averaged over five seeds. 
    \item \textbf{Own Coin Rate ($OCR_n$)}: Comparing the Own Coins Collected metric with the Total Coins Collected allows us to gauge the level of cooperation achieved, by determining:
    \begin{equation}
        OCR_n = \sum\limits_{i \in \{0,1\}} \sum\limits_{t=0}^{T-1} \frac{o_{t,i}}{c_{t,i}}
    \end{equation}
    This approach provides a detailed evaluation of the agents' behavior and their ability to cooperate within the Coin Game environment.
\end{enumerate}

\subsection{Training and Hyperparameters}
For all our experiments in the Coin Game environment, we train the agents over $\mu=200$ generations with a population size of $\eta=250$. Multiple executions have demonstrated that 200 generations are adequate for the evolutionary process to stabilize and produce an optimized agent. Furthermore, we accordingly choose the number of individuals in a generation to ensure diversity within the population to successfully explore the solution space and simultaneously not loose computational effort. The agents play against themselves in games of 50 steps, with each agent taking 25 steps. 

For the first part, evaluating a Multi-Agent setting in the Coin Game without changing the circuit architecture, the VQC is configured with 4 Variational Layers and $n_q=6$ qubits to embed the 36 features of the Coin Game, resulting in 76 parameters. After a preliminary study, we set the mutation power to $\sigma = 0.01$. We select the top $\tau=5$ agents to regenerate the following population. 

For the second part, investigating different architectural concepts, we adjusted some hyperparameters depending on the used concept to guarantee a good performance. The values were chosen after a short study. Other fundamental parameters are fixed throughout all experiments.

When evolution includes both recombination and mutation, the mutation rate is set to $\sigma=0.1$. For selection, the tournament size $\tau$ is set to 40\% of the population size $\mu$, which means $\tau=100$ for a population size of $\eta=250$. Again, we use $n_q=6$ qubits.

For the Layer-Based circuits, we set the initial layer count to 1, resulting in a total of 22 parameters, which includes 3 rotation angles per qubit and 4 bias values. Prototype-Based circuits begin with 8 layers, each consisting of 18 gates. Given that each gate in the gate set has an equal probability of being selected and 3 out of 4 gates are parameterized, the number of parameterized gates in such a circuit is approximately 108, resulting in an average of 112 parameters. A Gate-Based circuit starts with a total of 70 gates. Considering the probabilities of selecting parameterized gates, a Gate-Based circuit initially has about 52 parameterized gates, resulting in approximately 56 parameters at the beginning.

For Layer-Based and Prototype-Based circuits, the parameter mutation power is set to $\sigma_p=0.05$, while for Gate-Based circuits, it is set to $\sigma_p = 0.01$. The architecture mutation power $\sigma_a$ is set to 10 for Layer-Based and Prototype-Based circuits, and for Gate-Based circuits, it is set to $\sigma_a=1$. 

Each experiment is conducted using five different seeds (0 through 4) to ensure a more accurate performance assessment. Given the current limitations of quantum hardware, we use the Pennylane DefaultQubit simulator for all VQC executions. All experiments were run on nodes equipped with an Intel(R) Core(TM) i5-4570 CPU @ 3.20GHz.
\section{Results}
\label{sec: results}
In this section, we present the results of our experiments in the Coin Game environment. We tested all approaches with mutation and recombination combined and mutation only. For our first experiments, we additionally tested two classical neural networks. They consist out of two hidden layers, one with size 64 x 64 and the second one with size 3 x 4. We choose this size to gain better comparability of the model-size and performance ratios, because the number is close to the number of parameters we use in the VCQ approaches. Furthermore, we included agents in our tests that act randomly at every step to get a random baseline. For the second part of our experiments, we also started with a combination of mutation and recombination, as well as mutation only. A look at the results of these experiments, again led us to continue with the mutation-only strategy in the further experiments. We compare the
results of the Layer-Based, Gate-Based, and Prototype-Based circuits with each other, and additionally include a static baseline, where the circuit architecture remains unchanged throughout the generations.

\subsection{MAQRL in Coin Game}
\subsubsection{Comparing Generational Evolution Strategies} \label{sec: comp_GES}
Our goal is to assess the impact of different generational evolution strategies. First, we compare the performance of a mutation-only strategy (Mu) with two combined strategies involving both mutation and recombination. The first combined approach uses a random-point crossover recombination strategy (RaReMu), where crossover occurs at a randomly chosen point in the parameter vector. The second combined strategy employs a layer-wise crossover (LaReMu), selecting a random layer and applying the crossover after the last parameter of that layer. For all strategies, the mutation power $\sigma$ is fixed at 0.01.

\begin{figure}
    \centering
    \includegraphics[width=\linewidth]{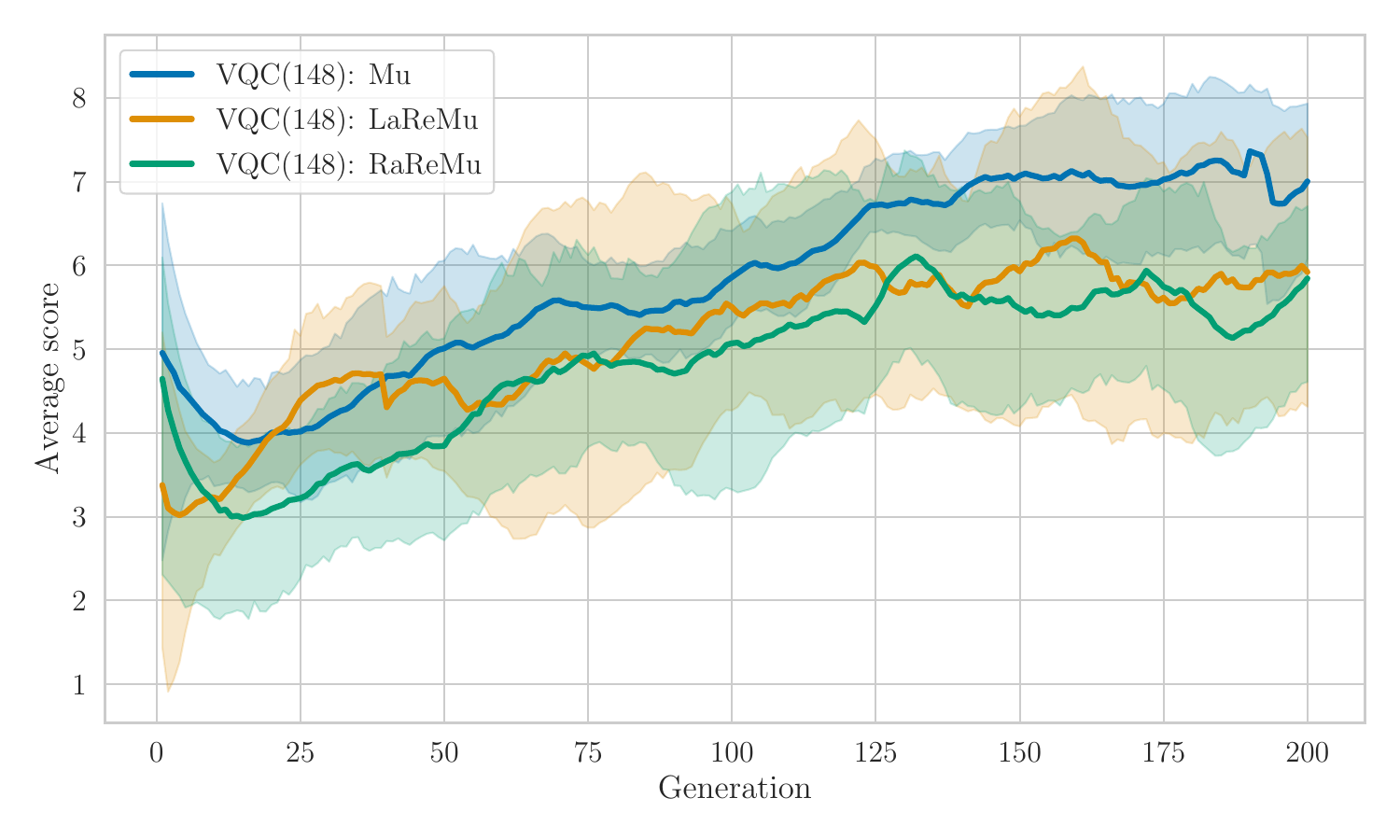}
    \caption{Average Score over the entire population. Each
individual has completed 50 steps in the Coin Game environment in each generation \cite{kolle2023multi}. }
    \label{fig: methods avg}
\end{figure}

\cref{fig: methods avg} illustrates that the mutation-only strategy outperforms the combined strategies. The mutation-only approach begins with an average reward of 5, dips slightly below 4 by the 17th generation, and then steadily increases, stabilizing around a score of 7 by the 140th generation. In contrast, the layer-wise recombination strategy starts at 3.3, rapidly ascends until the 30th generation, and then stabilizes, reaching an average reward of 6 by the 123rd generation with fluctuations thereafter. The random crossover strategy starts near the mutation-only strategy at 4.7, declines to 3 by the 17th generation, then climbs steadily until the 131st generation, reaching a score of 6 but eventually settling around 5.5, making it the least effective of the three methods.

\begin{figure}
     \centering
     \subfloat[][Total coins collected]{\includegraphics[width=0.32\textwidth]{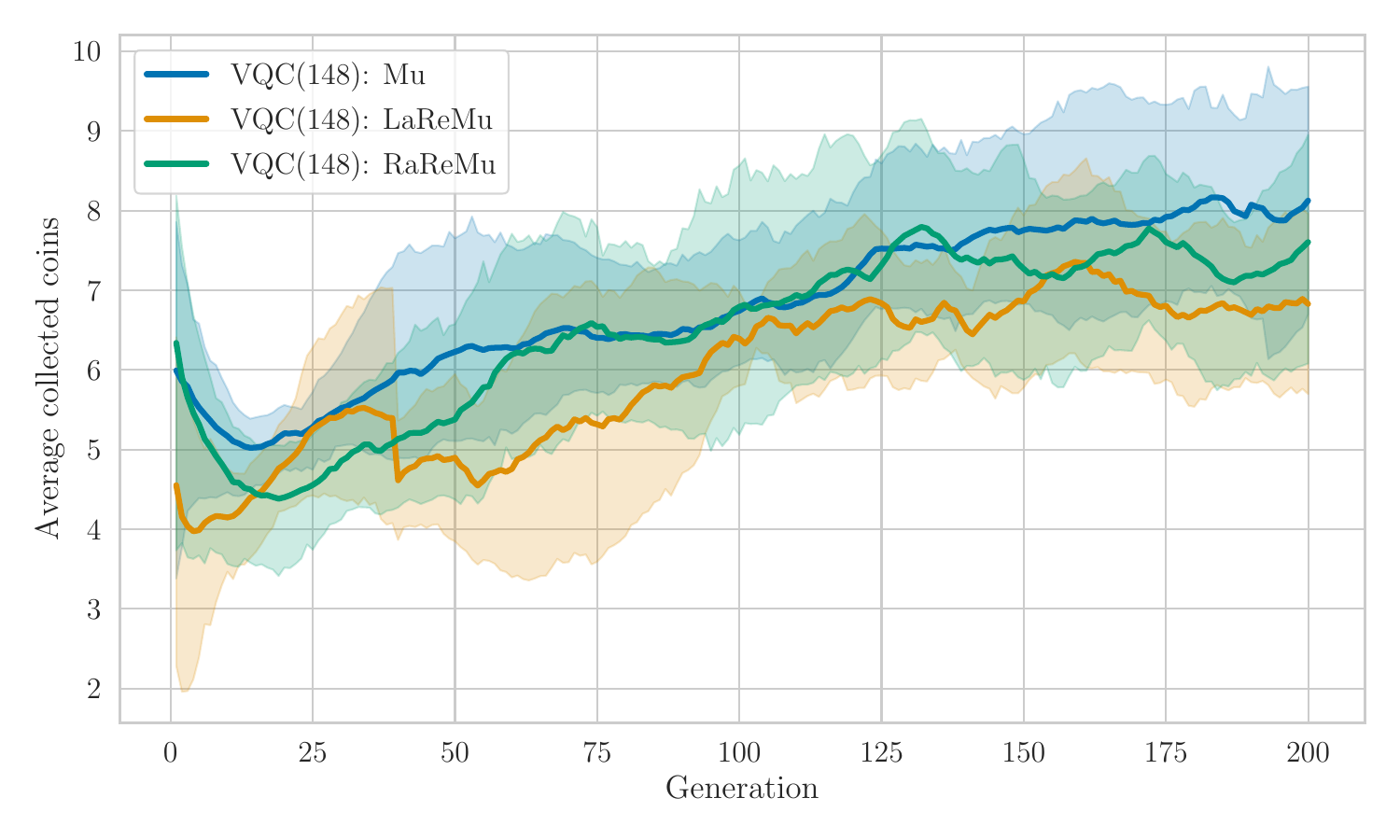}\label{fig:MethodsAvgCoins}}
     \subfloat[][Own coins collected]{\includegraphics[width=0.32\textwidth]{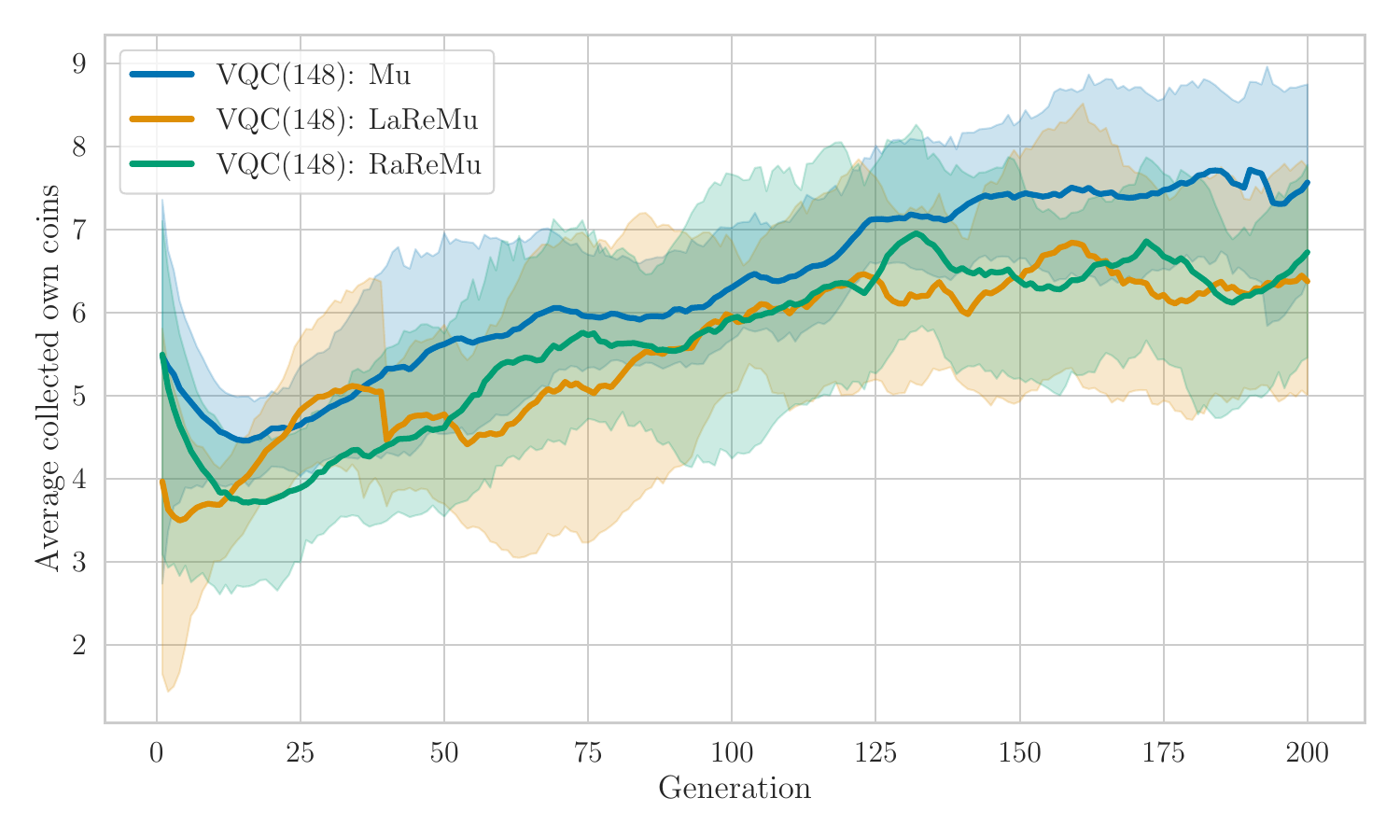}\label{fig:MethodsAvgownCoins}}
     \subfloat[][Own coin rate]{\includegraphics[width=0.32\textwidth]{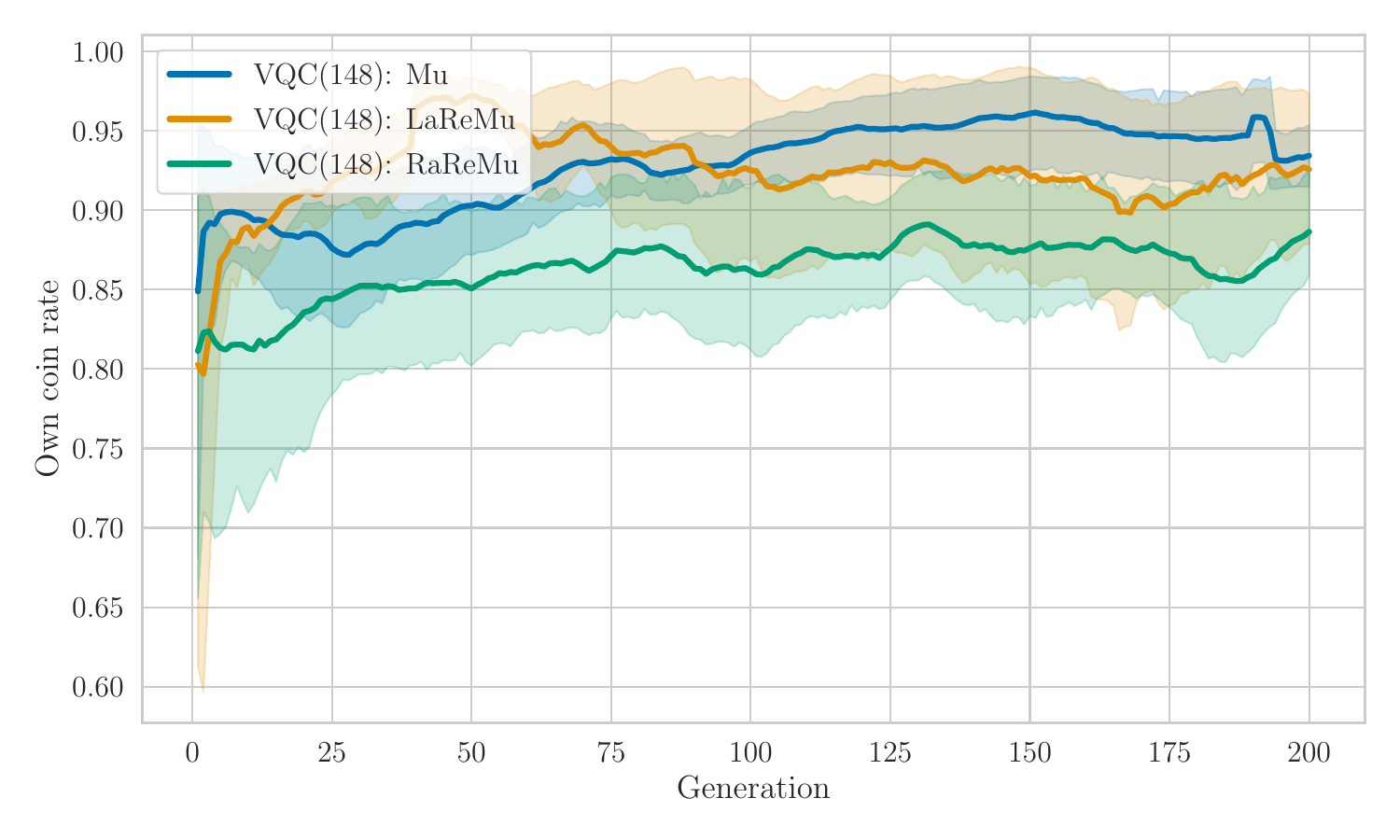}\label{fig:Methodsowncoinrate}}
     \caption{Comparison of (a) average coins collected, (b) average own coins collected and the own coin rate (c) in a 50 step Coin Game each generation, averaged over 10 seeds \cite{kolle2023multi}.}
     \label{fig:Methodscoins_comp}
\end{figure}

We also evaluated the average number of coins collected during the experiments. The mutation-only strategy consistently collects more coins, as shown in \cref{fig:MethodsAvgCoins}. While there are periods where all strategies have nearly identical coin counts, the mutation-only strategy generally leads, with up to a 2-coin gap at times. This is reflected in the coin rate (\cref{fig:Methodsowncoinrate}), where the mutation-only strategy shows a steady increase over generations, indicating enhanced agent cooperation in the testing environment.

The comparison of both strategies lead to the result, that the mutation-only strategy consistently achieves the highest coin rate and collects the most coins, aligning well with our goal of maximizing the reward. The layer-wise recombination strategy shows an initial surge in performance, correlating with an increase in collected coins and coin rate, but its advantage diminishes after the 90th generation. For the random crossover strategy we observe that it collects more coins than the layer-wise approach, but has a significantly lower coin rate, resulting in lower overall rewards.

All in all, the mutation-only strategy outperforms the combined strategies in our experiments, achieving the highest rewards and is best aligning with our objective of maximizing reward. Consequently, subsequent experiments will exclusively utilize the mutation-only approach for the VQCs.

\subsubsection{Evaluating VQC Performance with Different Layer Counts}
We investigate the performance dynamics of VQCs with varying layer counts: specifically, 4, 6, 8, and 16 layers. The relationship between layer counts and parameters is governed by the formula $3*n*6+4$, where $n$ is the number of layers, Therefore, VQCs with 4, 6, 8, and 16 layers utilize 76, 112, 148, and 292 parameters respectively. All VQCs were trained using the mutation-only approach, with the mutation strength set to $\sigma=0.01$.

\begin{figure}[ht]
    \centering
    \includegraphics[width=\linewidth]{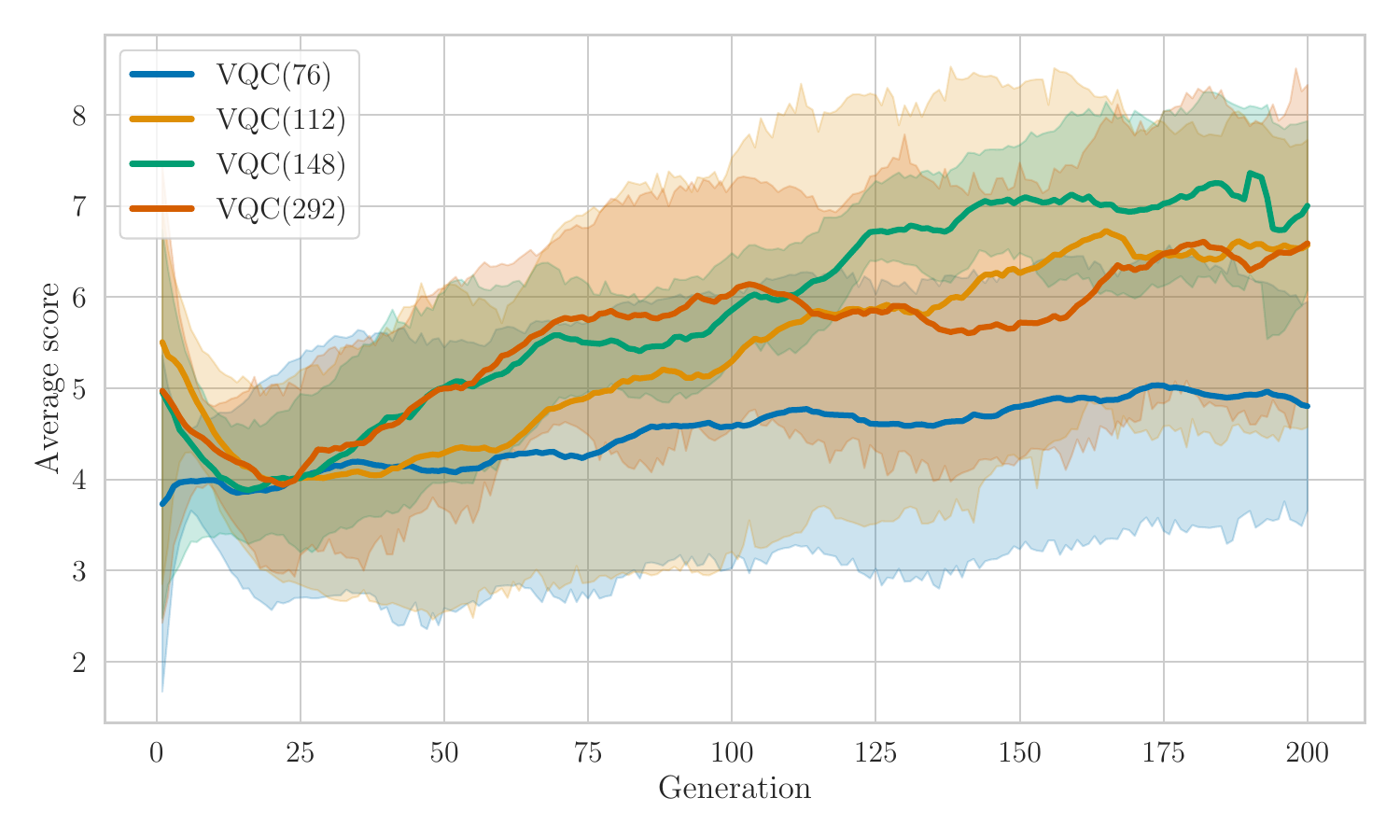}
    \caption{Average Score over the entire population. Each individual has completed 50 steps in the Coin Game environment each generation. \cite{kolle2023multi}}
    \label{fig:LayersAvgScore}
\end{figure}

\cref{fig:LayersAvgScore} shows that, except for the 4-layer VQC, which starts slightly below 3, all VQCs begin with scores between 5 and 5.5. Each VQC stabilizes around a reward of 4 by the 25th generation. The 4-layer VQC gradually increases, maintaining an average reward of 5 from the 175th generation onward. For the 6-layer VQC we observe a steady rise until the 62nd generation, with a more pronounced increase thereafter, reaching a maximum at 6.7 around the 165th generation and then oscillating around 6.5. The 8-layer VQC consistently outperforms the others, reaching a reward of 5.5 by the 70th generation, then climbing to 7 by the 140th generation after a brief plateau. The 16-layer VQC displays significant growth between the 25th and 70th generations, stabilizing around 6 before rising again to 6.5 around the 160th generation.

\begin{figure}
     \centering
     \subfloat[][Total coins collected]{\includegraphics[width=0.32\textwidth]{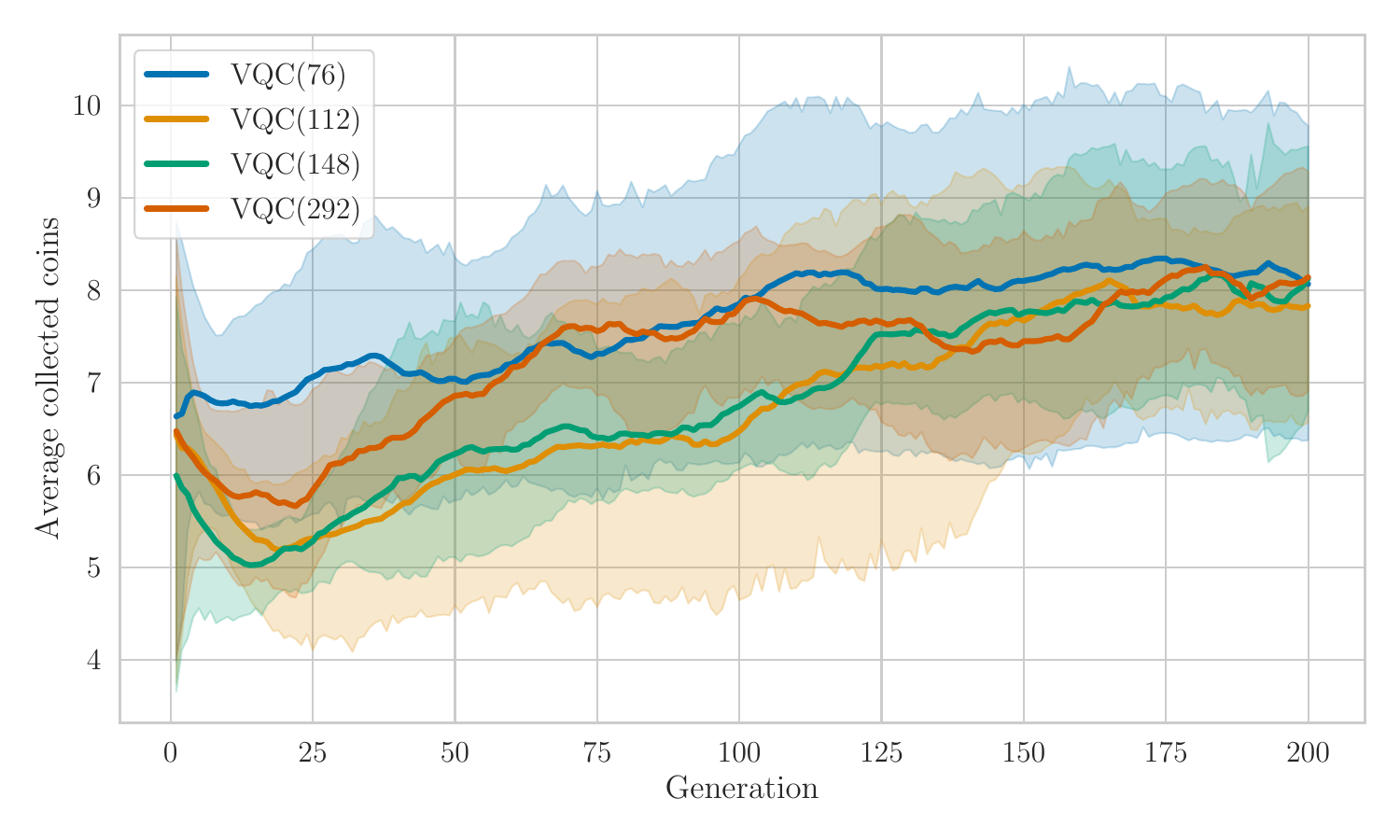}\label{fig:LayersAvgCoins}}
     \subfloat[][Own coins collected]{\includegraphics[width=0.32\textwidth]{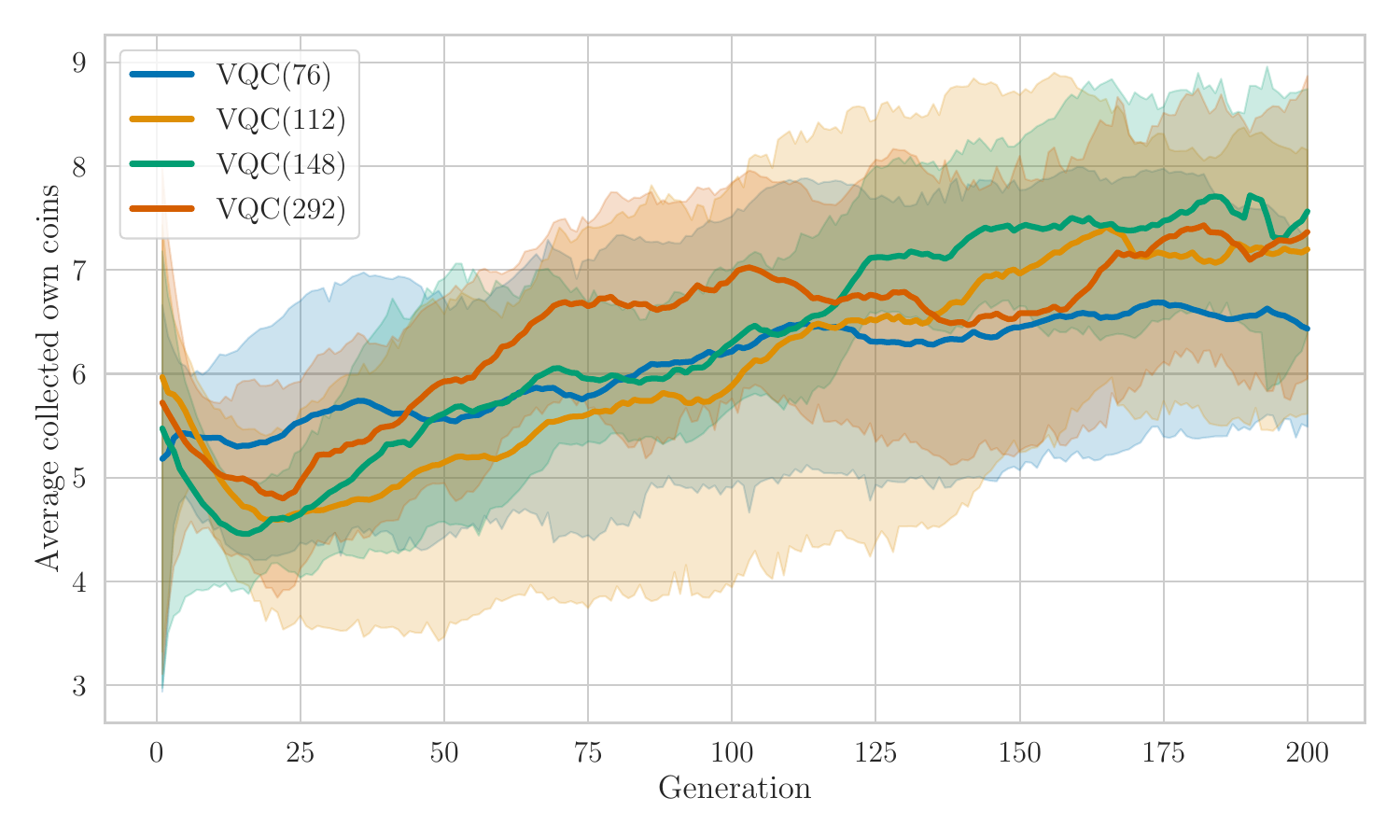}\label{fig:LayersAvgownCoins}}
     \subfloat[][Own coin rate]{\includegraphics[width=0.32\textwidth]{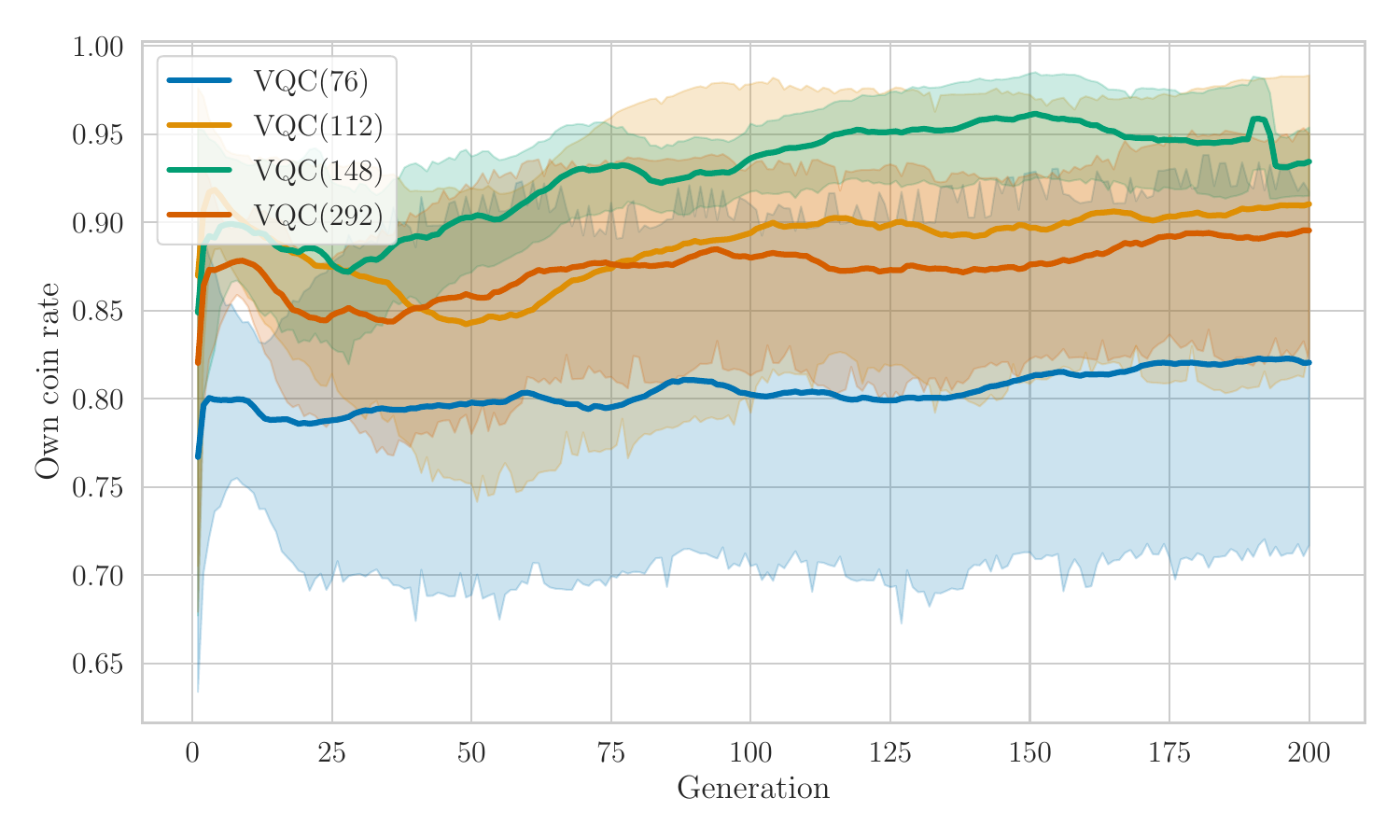}\label{fig:Layersowncoinrate}}
     \caption{Comparison of (a) average coins collected, (b) average own coins collected and the own coin rate (c) in a 50 step Coin Game each generation, averaged over 10 seeds. \cite{kolle2023multi}}
     \label{fig:Layercoins_comp}
\end{figure}

Next, we next examine the average coin collection in \cref{fig:LayersAvgCoins} to provide a comprehensive understanding. The 4-layer VQC consistently leads the coin collection metric, increasing from just below 7 to 8. The 6-layer VQC starts at 6.5, dips to 5.2 by the 23rd generation, and then rises to 8 by the 165th generation. Despite achieving the highest average reward, the 8-layer VQC begins at 6 coins and only stabilizes around 8 coins after the 180th generation. The 16-layer VQC, rapidly increases from the 24th to the 103rd generation, briefly declines, and then fluctuates around 8 coins. By the later generations, VQCs with more than 4 layers converge to collect approximately 8 coins each.

\cref{fig:Layersowncoinrate} reveals for the own coin count that, except for the 4-layer VQC, all VQCs initially decline before ascending. The 4-layer VQC shows a modest but steady rise, ending with a count of 6.5 coins. The 6-layer VQC is slowest to begin its ascent, eventually stabilizing around 7.2 coins. The 8-layer VQC starts rising earlier, reaching about 7.5 coins by the end. The 16-layer VQC, known for its rapid early rise, maintains around 7 coins after the 100th generation. The 4-layer VQC lags, collecting more than one own coin fewer than the other VQCs.

When focusing on the own coin rate, the 4-layer VQC performs the worst. The 6-layer and 16-layer VQCs show similar performance patterns in terms of own coin rate and overall reward. The 8-layer VQC stands out with the highest own coin rate and comparable coin count, achieving the highest reward.

Our analysis identifies the 8-layer VQC as the top performer. As a result, we will use this configuration, combined with the optimal evolutionary strategy outlined in \cref{sec: comp_GES}, for further experiments. This analysis highlights that a higher layer count does not necessarily lead to better performance, as the 16-layer VQC did not match the achievements of the 8-layer VQC. However, the experiments do not conclusively establish the performance dynamics beyond 200 generations.

\subsubsection{Comparing Quantum and Classical Approaches}
\begin{figure}[ht]
    \centering
    \includegraphics[width=\linewidth]{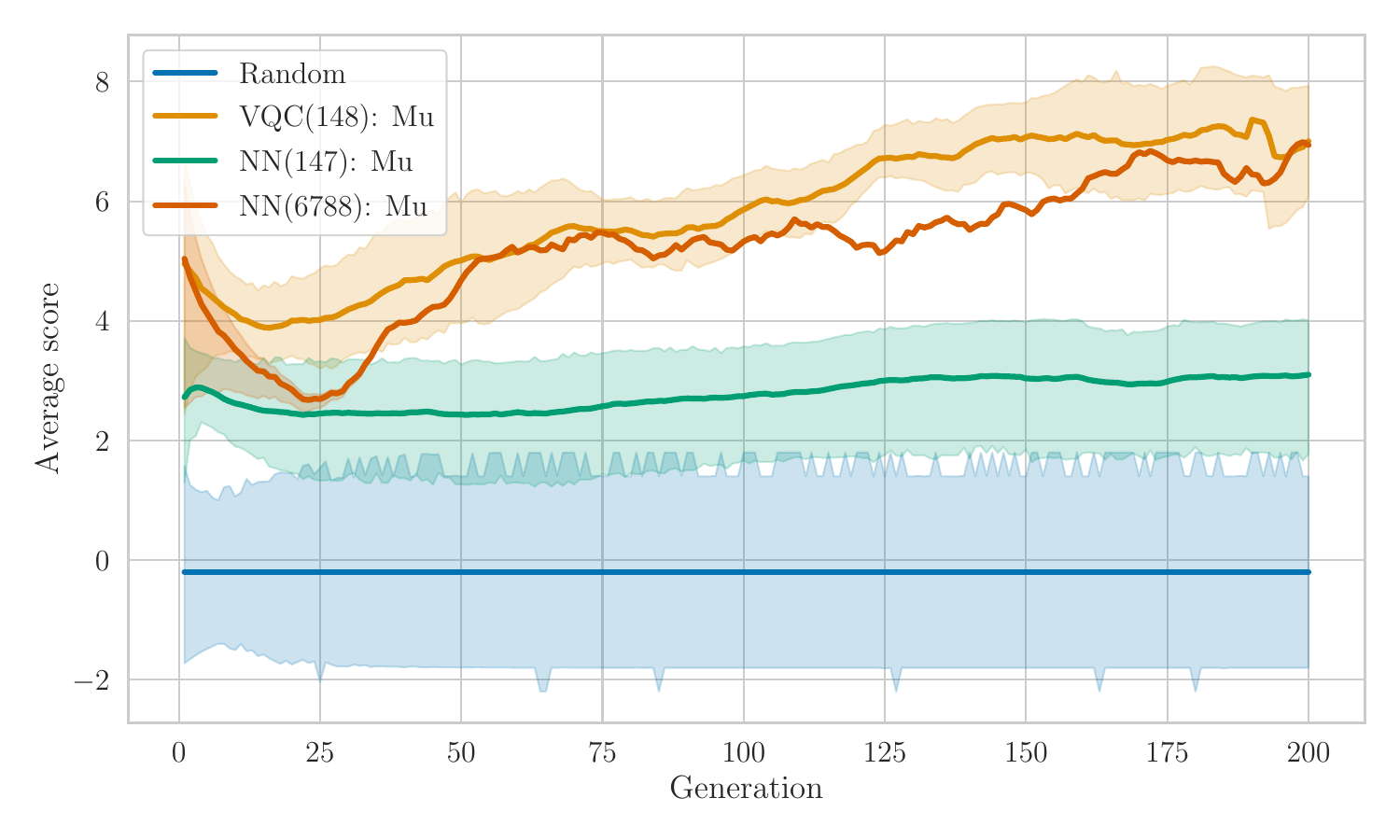}
    \caption{Average Score over the entire population. Each individual has completed 50 steps in the Coin Game environment each generation.}
    \label{fig:BestAvgScore}
\end{figure}

\begin{figure}
     \centering
     \subfloat[][Total coins collected]{\includegraphics[width=0.32\textwidth]{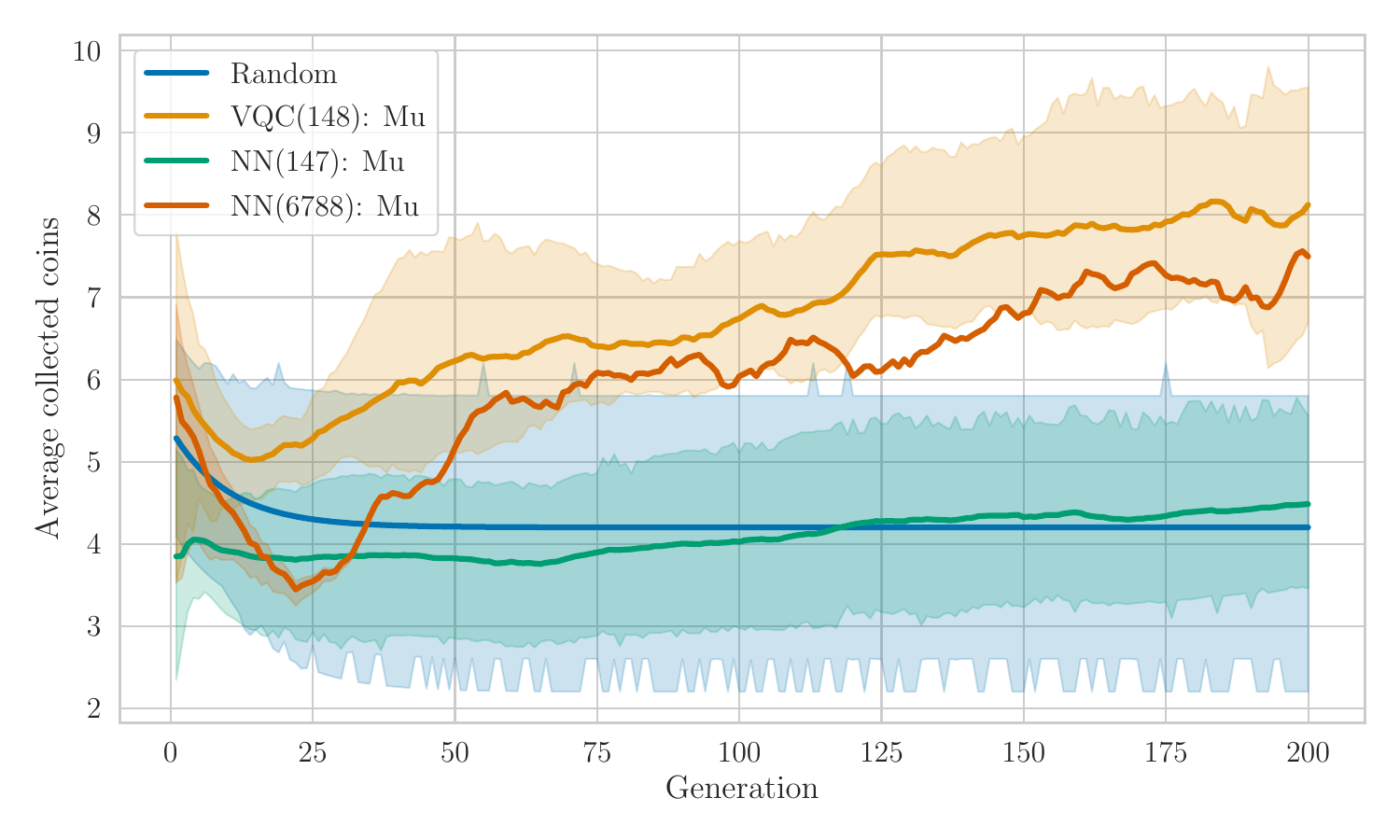}\label{fig:BestAvgCoins}}
     \subfloat[][Own coins collected]{\includegraphics[width=0.32\textwidth]{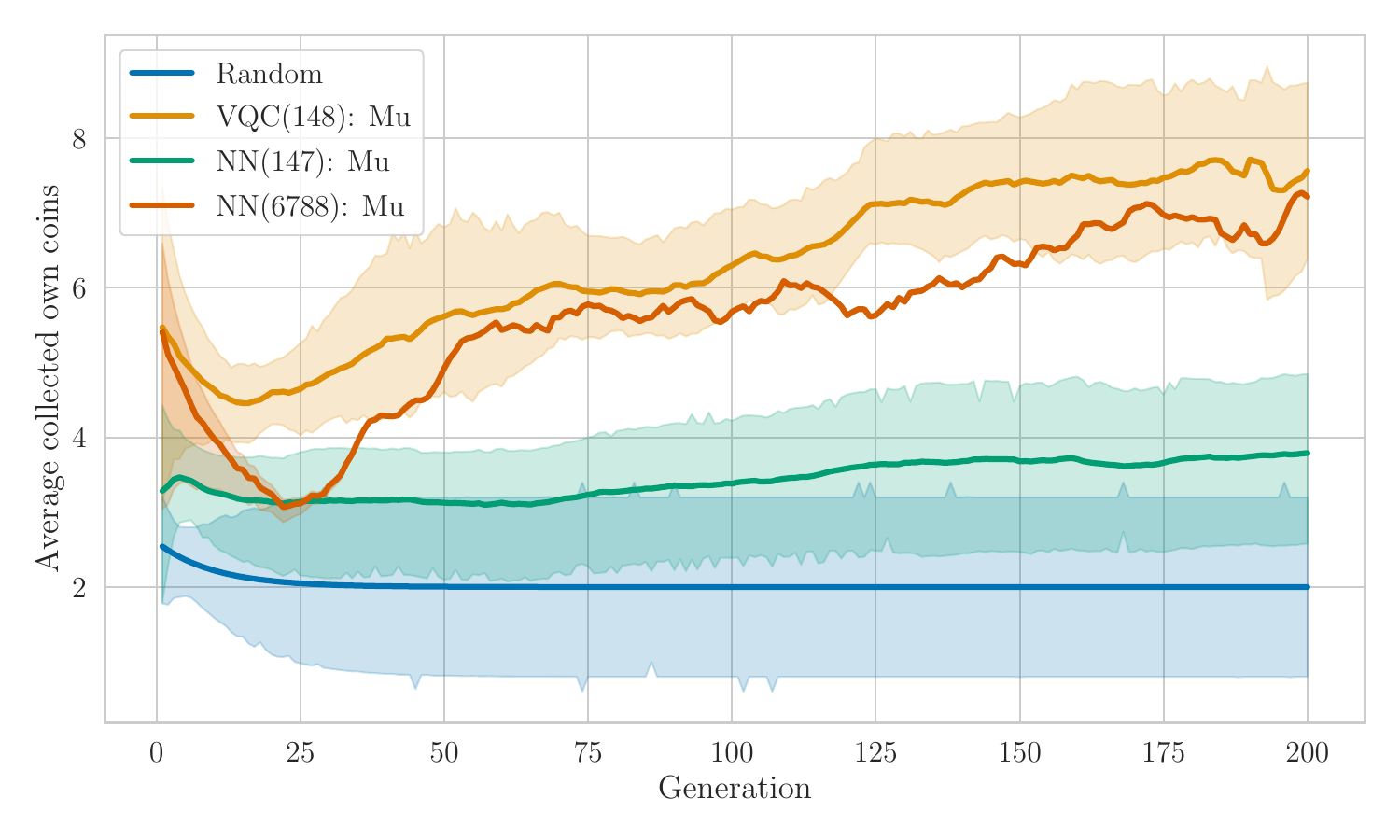}\label{fig:BestAvgownCoins}}
     \subfloat[][Own coin rate]{\includegraphics[width=0.32\textwidth]{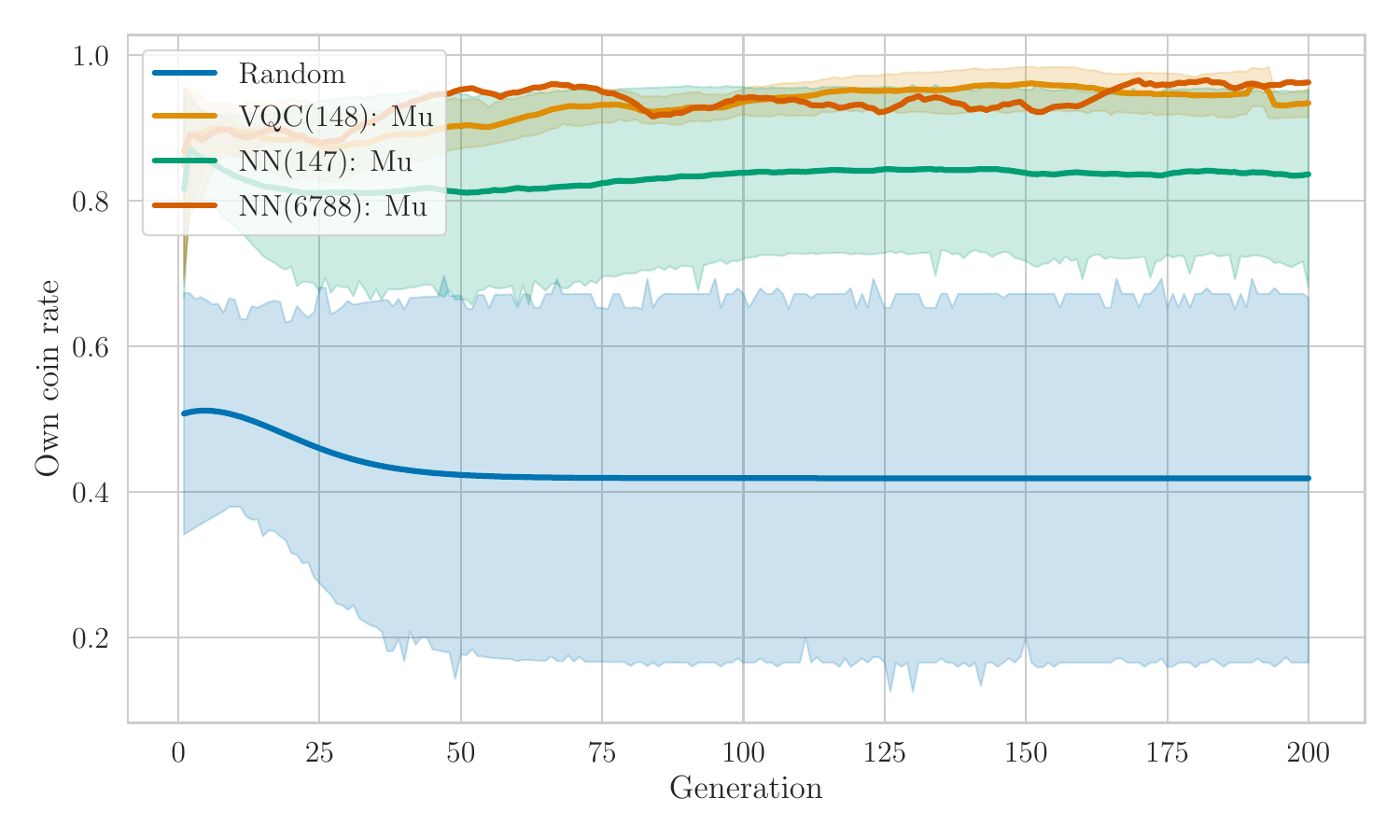}\label{fig:Bestowncoinrate}}
     \caption{Comparison of (a) average coins collected, (b) average own coins collected and the own coin rate (c) in a 50 step Coin Game each generation, averaged over 10 seeds.}
     \label{fig:Bestcoins_comp}
\end{figure}

\textbf{Comparative Analysis of VQC Approaches vs. Random Baseline}
We first evaluate the performance of our VQC methods compared to a random baseline. As shown in \cref{fig:BestAvgScore}, the random agents' score hovers around zero, reflecting the zero-sum nature of the cooperative sequential coin game. In contrast, VQC agents trained through evolutionary methods significantly outperform random agents, achieving an average score around 7. Figure \cref{fig:BestAvgCoins} illustrates that VQC agents effectively learn to collect coins, unlike random agents. The number of own coins collected, as shown in \cref{fig:BestAvgownCoins}, aligns with the overall coins collected, indicating that VQC agents consistently outperform the random baseline. Interestingly, \cref{fig:Bestowncoinrate} reveals that cooperation does not increase over time for either approach. In summary, the trained VQC agents exhibit superior performance across all metrics, confirming the efficacy of the training process.

\textbf{VQC vs. Small Neural Network}
With VQC methods proving superior to random agents, we next compare their performance against a small neural network with a comparable parameter count. Following Chen et al. \cite{chen2022variational}, we recognize that VQCs possess greater expressive power than traditional neural networks, defined by their capacity to represent complex functions with fewer parameters. Our VQC has 148 parameters (3 * 6 * 8 + 4), while the neural network, composed of two hidden layers with dimensions 3 and 4, has 147 parameters. Both models are trained using mutation only, with a mutation power $\sigma = 0.01$.

In \cref{fig:BestAvgScore}, the neural network's rewards fluctuate between 2.5 and 3, while the VQC, despite a slower initial learning curve, achieves a significantly higher score. The small neural network’s limited number of hidden units likely accounts for its poorer performance. \cref{fig:BestAvgCoins} demonstrates that the neural network collects fewer coins than random agents until generation 115, after which it only slightly surpasses their performance. In comparison, the VQC consistently collects about twice as many coins. As shown in \cref{fig:BestAvgownCoins}, the neural network is able to outperform random agents in terms of own coins collected, leading to a better own coin rate (\cref{fig:Bestowncoinrate}). However, the VQC still significantly outperforms the neural network with nearly the same parameter count, highlighting the VQC's superior capabilities in this environment.

\textbf{VQC vs. Larger Neural Network} 
Given the superior performance of VQCs with a similar parameter count, we next compare the VQC to a significantly larger neural network. Both models continue using mutation-only evolution with a mutation power of $\sigma = 0.01$. The larger neural network features two hidden layers of size 64, resulting in 6788 parameters.

As depicted in \cref{fig:BestAvgScore}, the VQC and the larger neural network achieve similar results over time, with minor differences in average score from generation 50 onwards, both stabilizing around a score of 7. Despite having 46 times more parameters, the larger neural network only marginally outperforms the VQC initially but converges to a similar performance level in the long run. \cref{fig:BestAvgCoins} shows that the VQC starts with a higher initial coin collection, while the neural network compensates with a steeper learning curve, ultimately achieving a comparable number of coins. In terms of own coins collected, shown in \cref{fig:BestAvgownCoins}, both models perform similarly. Initially, the neural network achieves a higher own coin rate around generation 25, but the VQC maintains a slight edge between generations 80 and 162, with the neural network slightly outperforming the VQC towards the end (\cref{fig:Bestowncoinrate}).

In conclusion, despite the larger neural network having 46 times more parameters, the performance difference between the two models is minimal. This demonstrates that VQCs can achieve comparable performance with a 97.88\% reduction in parameters, supporting the findings of Chen et al. \cite{chen2022variational} regarding the expressive power of VQCs. 

\subsection{Architectural Variations}
In the second part of our results we examine how different architecture strategies during the evolution process influence the performance of the agents. Therefore, we investigate the efficiency of recombination and mutation, and we compare the Gate-Based, Layer-Based and Prototype-Based approaches against each other looking at the score, coin and gate count metrics. We conclude the results section with a look at the performance of a static baseline (i.e. an unchanged circuit architecture throughout the whole evolution process) compared to architectural evolution. 

\subsubsection{Assessment of Evolutionary Strategies: ReMu vs. Mu}

In this section, we evaluate the effectiveness of two evolutionary strategies for generating successive generations: recombination combined with mutation (ReMu) and mutation-only (Mu). The goal is to understand the performance of these strategies to optimize the evolutionary process in future experiments. As the performance trends of recombination versus mutation-only are consistent across different approaches, in our analysis we focus on the Gate-Based approach.

\begin{figure}[ht]
    \centering
    \includegraphics[width=\linewidth]{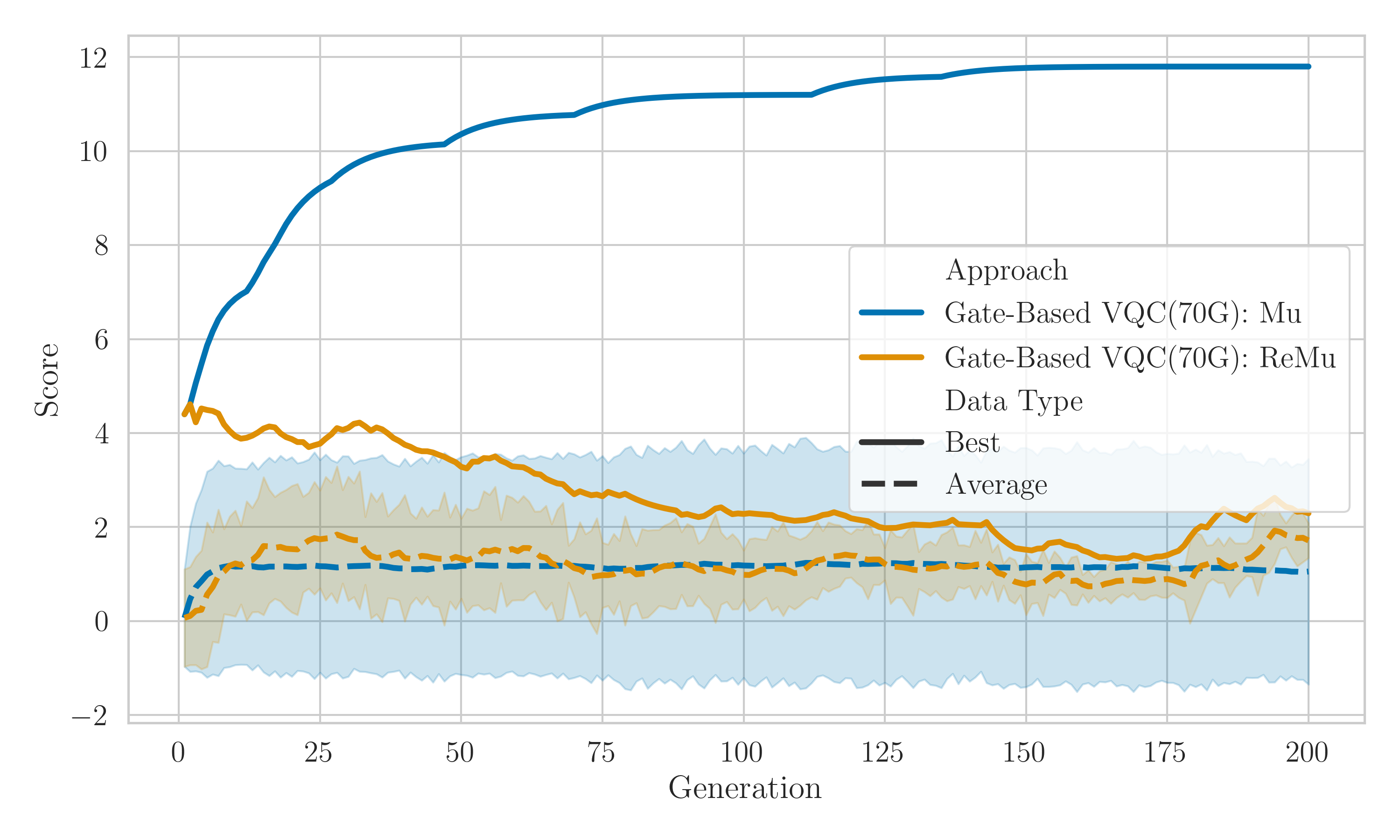}
    \caption{Best and average score for the different evolutionary algorithm methods averaged over 5 seeds.}
    \label{fig:Score_ReMuMu}
\end{figure}

Our findings reveal a significant difference in the performance of the ReMu and Mu strategies. As illustrated in \cref{fig:Score_ReMuMu}, the recombination approach failed to yield good solution candidates. The best score achieved by the ReMu strategy declined from 4.5 to 1.3 by the 170th generation, with a minor recovery to just above 2, resulting in an overall decline of more than 2 points. In contrast, the best run of the mutation-only approach showed a steady and consistent improvement, with scores rising from 4.5 to almost 12 by generation 150.

For the ReMu strategy, the gap between the average score and the best score diminished over time, indicating a convergence towards lower-quality solutions. By the 125th generation, the best score was only about 0.6 points above the average. This suggests that the recombination method might introduce complexities or incompatibilities that hinder the evolutionary process, leading to suboptimal solutions and reduced diversity.

In contrast, the consistent upward trajectory in performance of the mutation-only strategy indicates a more efficient navigation of the solution space, leading to the discovery of higher-quality solutions.

Based on these findings, we conclude that the mutation-only strategy is more effective for our purposes. The ReMu strategy's introduction of complexities appears to hinder the evolutionary process, leading to less optimal solutions. Therefore, we will exclusively employ the mutation-only strategy in future experiments to maximize the potential for discovering optimal solutions.

\subsubsection{Comparative Analysis of Architectural Evolution Strategies}
Now, we evaluate the performance of three Variational Quantum Circuit architecture structures and their respective evolutionary strategies: Layer-Based, Gate-Based, and Prototype-Based. Each approach presents a unique design for the variational layers and employs distinct architectural changes within the Evolutionary Algorithm (EA). 

\begin{figure}[ht]
    \centering
    \includegraphics[width=\linewidth]{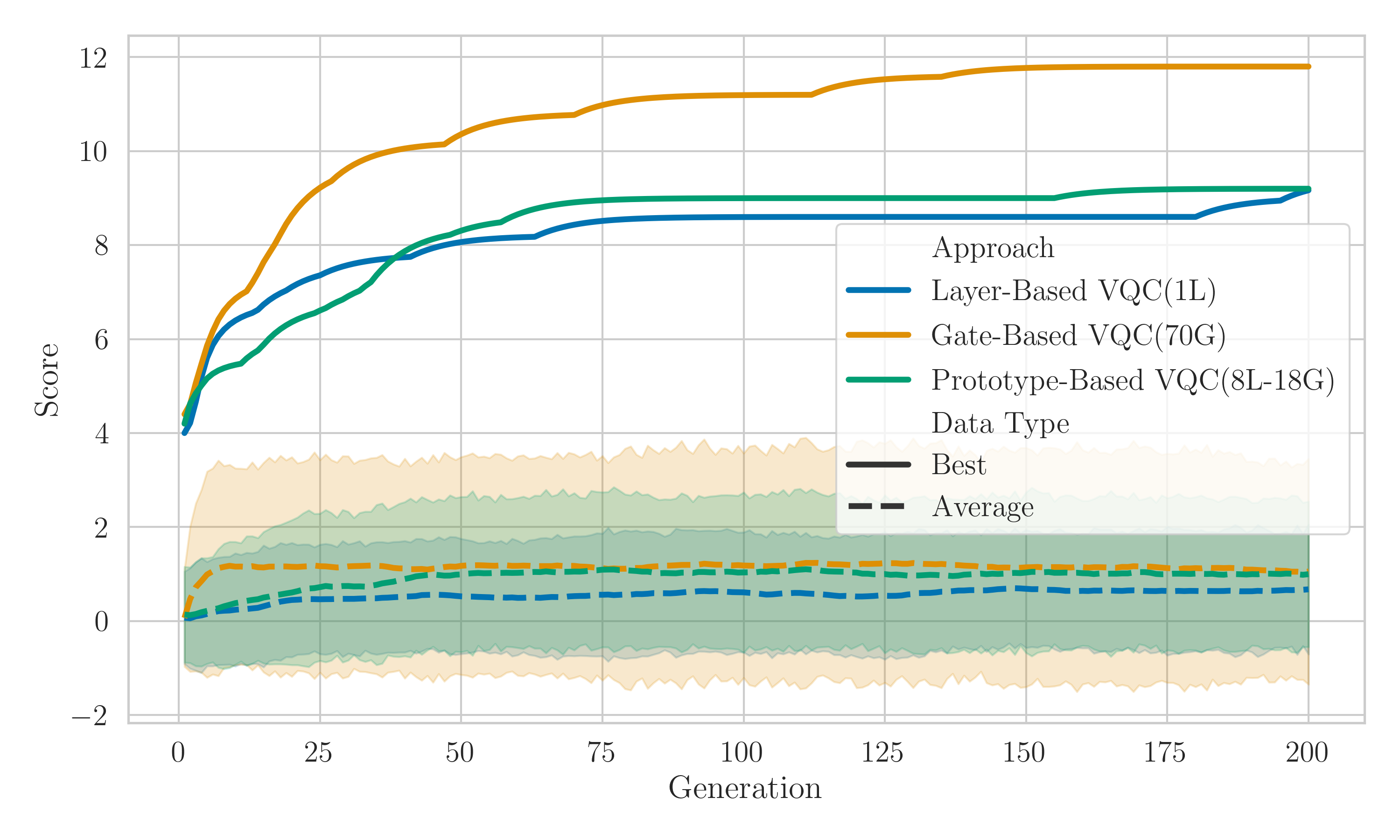}
    \caption{Best and average score for the different architectural approaches averaged over 5 seeds.}
    \label{fig:Score_Methods}
\end{figure}

\cref{fig:Score_Methods} illustrates the evolution of the score for all three approaches. Initially, the average scores start at 0. The Layer-Based and Prototype-Based approach stabilize around 1 by generation 25. Notably, the Gate-Based approach reaches this stabilization point faster, achieving it by the 6th generation and consistently maintaining a higher score than the other two approaches. The average score of the Layer-Based approach, while consistently lower, differs only slightly from the others.

When examining the best scores, the differences become more pronounced. All approaches begin with a best score around 4. However, the Gate-Based approach rapidly ascends, reaching a best score of 10 by generation 35, which already surpasses the final best scores (within 200 generations) of the Layer-Based and Prototype-Based approaches. The Gate-Based approach continues to improve, reaching nearly 12 by generation 155. In contrast, the Layer-Based and Prototype-Based approaches peak around 9 by generation 75 and show little change afterward.

\begin{figure} [ht]
     \centering
     \subfloat[][Total coins collected]{\includegraphics[width=0.32\textwidth]{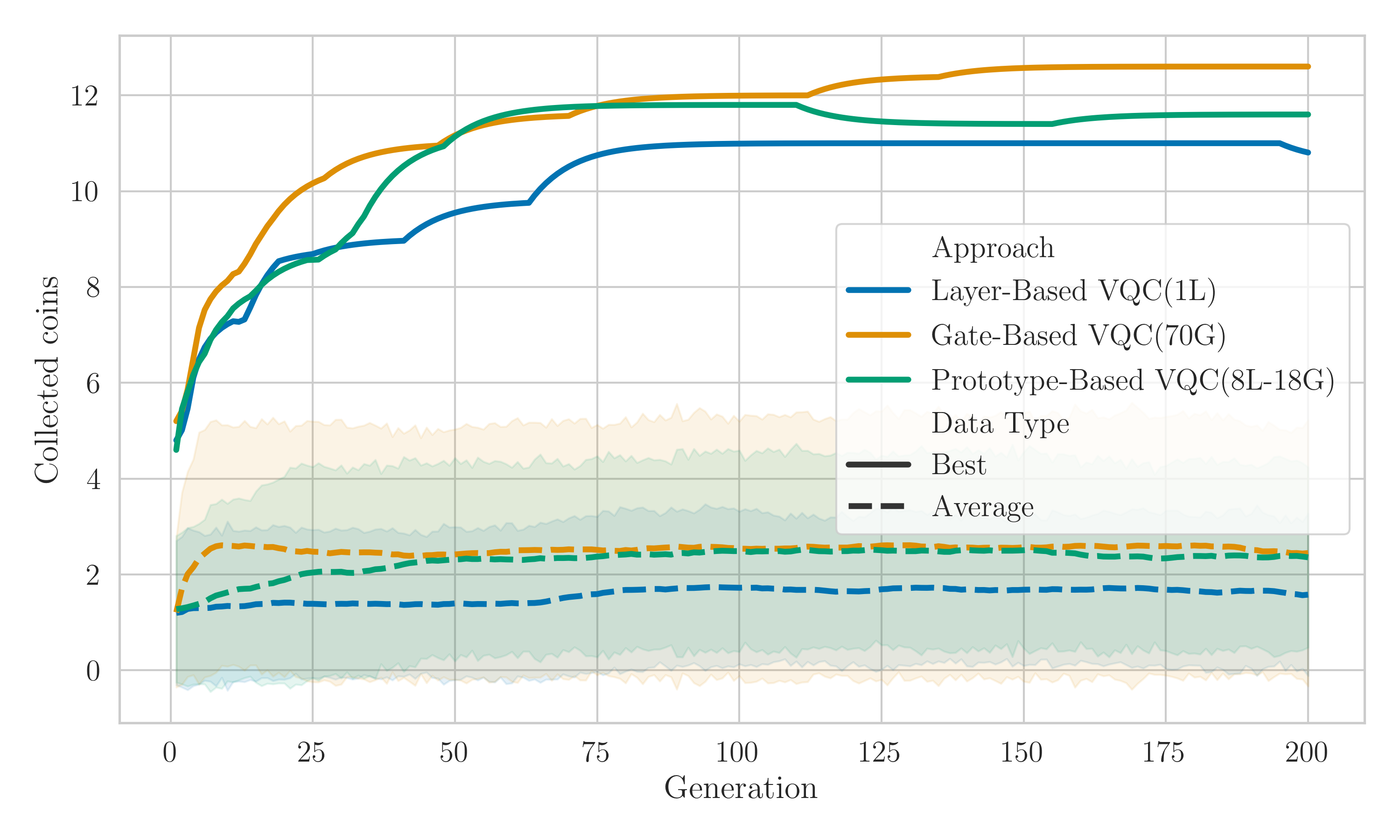}\label{fig:Methods_totalCoins}}
     \subfloat[][Own coins collected]{\includegraphics[width=0.32\textwidth]{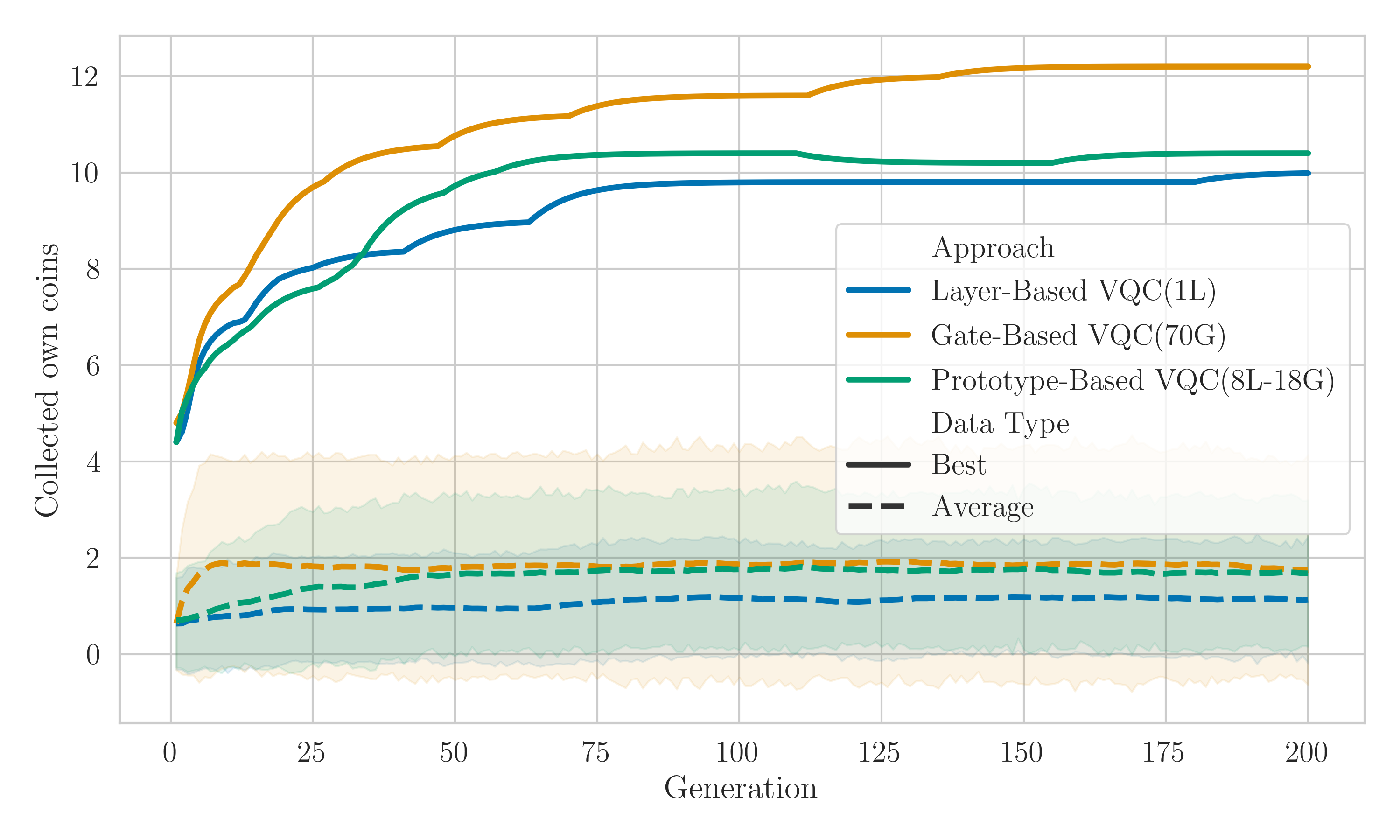}\label{fig:Methods_ownCoins}}
     \subfloat[][Own coin rate]{\includegraphics[width=0.32\textwidth]{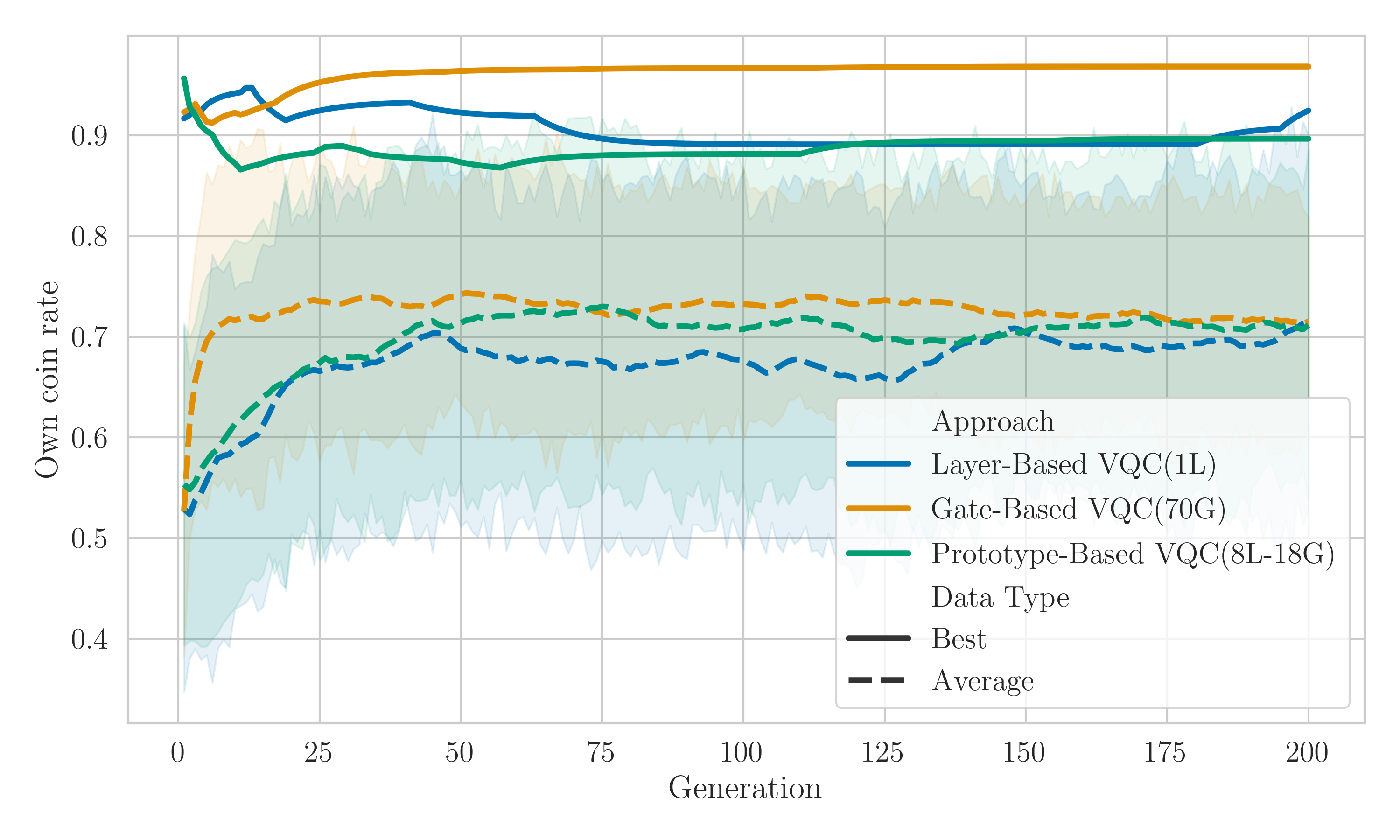}\label{fig:Methods_ownCoinRate}}
     \caption{Comparison of best and average (a) total coins collected, (b) own coins collected, and (c) own coin rate for the different architectural approaches averaged over 5 seeds.}
     \label{fig:Methods_Coins_comp}
\end{figure}

To gain further insight into the scores, we also analyze the total coins collected and the own coin rate, which represents the percentage of collected coins belonging to the player. \cref{fig:Methods_totalCoins} shows that, when looking at the best performing agents, the Gate-Based and Prototype-Based approaches perform similarly in terms of total collected coins until generation 110, both reaching up to 12 coins. After this point, the Gate-Based approach increases to 12.6 coins, while the Prototype-Based approach declines to 11.7. The Layer-Based approach, on the other hand, collects fewer coins than the Prototype-Based approach after the 30th generation, plateauing at 11 coins by generation 90.

The own coin rate of the best agents (\cref{fig:Methods_ownCoinRate}) reveals that the Prototype-Based approach starts strong at 0.95 but drops to 0.87 by the 10th generation, stabilizing at 0.89 by generation 115. The Layer-Based approach performs better initially, reaching 0.95 by the 12th generation, but falls below 0.9, climbing back to 0.93 by the 200th generation. The Gate-Based approach, however, consistently outperforms the others starting from generation 17, achieving a rate of 0.97 by generation 60. The average own coin rate for all approaches increases from 0.53 to 0.7, with the Gate-Based approach showing the steepest initial increase.

These results indicate that the superior own coin rate of the Gate-Based approach, along with a higher total coin collection, contributes to its higher overall score. This suggests that the Gate-Based approach promotes a more cooperative play style, leading to better performance in the Coin Game. The flexibility of the Gate-Based architecture may enable more effective adaptation to the task compared to the more restrictive Layer-Based and Prototype-Based architectures.

\begin{figure}[ht]
    \centering
    \includegraphics[width=\linewidth]{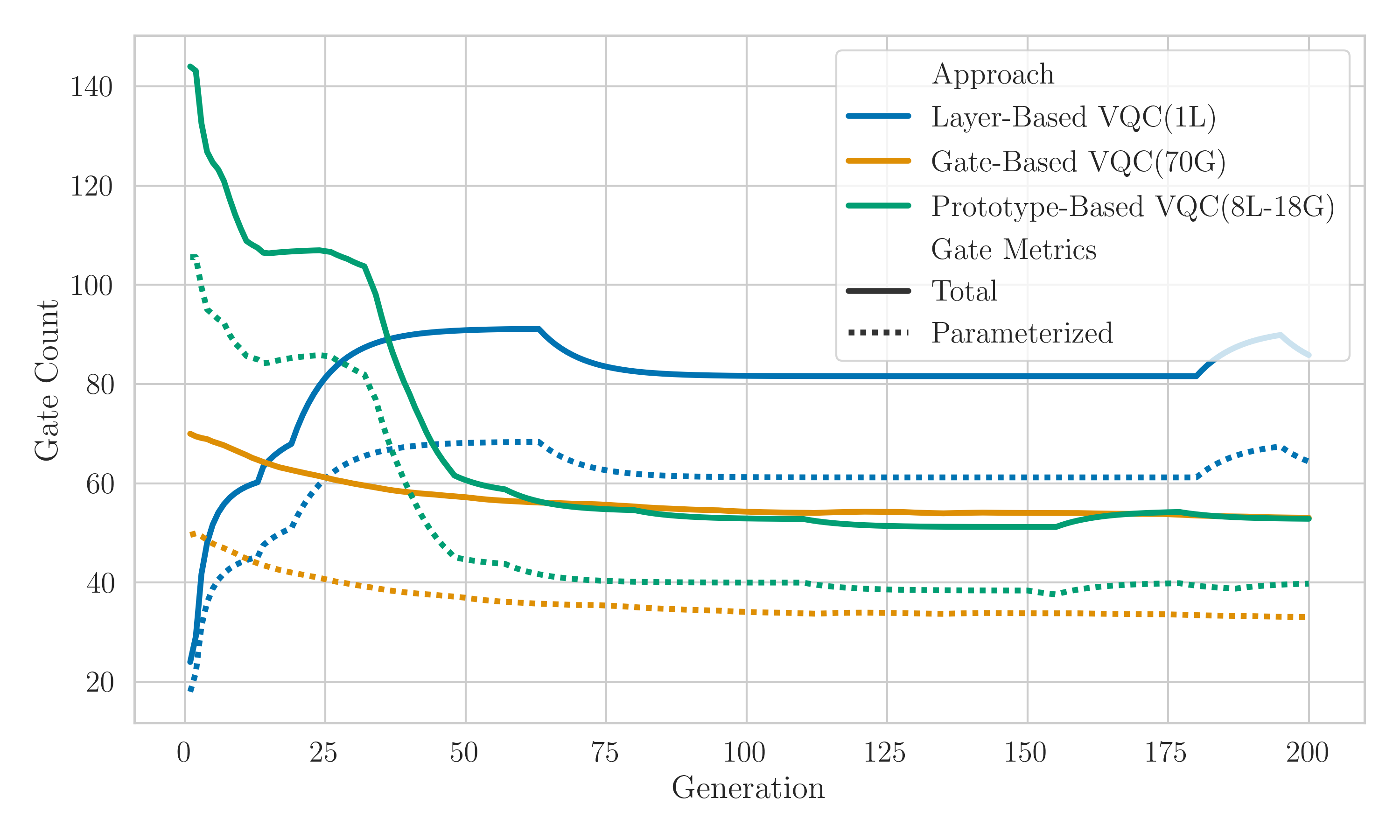}
    \caption{Evolution of total and parameterized gate count for the different architectural approaches averaged over 5 seeds.}
    \label{fig:Gate_count}
\end{figure}

To examine how circuit size evolves, \cref{fig:Gate_count} analyzes the total and parameterized gate counts of the best circuits for each approach: For the Gate-Based approach the total gate count decreases from 70 to 53, and parameterized gates from 50 to 33. In contrast, the total gate count for the Prototype-Based approach drops significantly from 144 to 53, and parameterized gates from 106 to 40. The lowest number of initial gates shows the Layer-Based approach: Starting with 22 total gates and 18 parameterized gates, it rapidly increases to about 90 total and 68 parameterized gates by generation 60, then slightly decreases to 80 and 60 gates by generation 80, before rising again by generation 180.

The Gate-Based and Prototype-Based approaches, which allow more flexibility in gate composition, both settle around 53 total gates and 35 parameterized gates. This suggests that a higher gate count does not necessarily correlate with better performance. The results highlight that a more adaptable architecture, like the Gate-Based approach, can outperform a larger, less flexible one. In the current NISQ era, minimizing circuit sizes is crucial to reduce errors and noise and to benefit from shorter computation times. The ability of the algorithm to maximize scores while reducing the number of parameterized gates accelerates the evolutionary process, as fewer parameters require less time to evolve.

The rapid improvement and superior performance of the Gate-Based approach indicate that its flexible architecture provides a significant advantage in exploring and optimizing the solution space. This approach's adaptability allows for more dynamic changes, leading to better solution candidates. In contrast, the Layer-Based and Prototype-Based approaches, while showing improvements, are constrained by their less flexible architectures, resulting in lower scores and slower progress. In summary, the Gate-Based approach, with its less restrictive architecture, proves to be crucial for achieving higher performance and faster convergence in VQCs. It outperforms the more constrained Layer-Based and Prototype-Based approaches in both score and computational efficiency.

\subsubsection{Comparison of Gate-Based VQC and Static Baseline}

Given the superior results of the Gate-Based approach in previous experiments, we compare it to a static baseline to evaluate its efficiency and performance. We analyzed two versions of the Gate-Based approach: one with 70 initial gates and a scaled-down version with 50 initial gates. 

The static baseline consists of 8 layers of the Layer-Based VQC, with only parameter mutations ($\sigma_p = 0.01$) applied, and generates the next generation from the top 5 agents. The Gate-Based VQC uses a mutation power of $\sigma_p = 0.01$ for parameters and $\sigma_a = 1$ for architecture, employing 40\% of the population for tournament selection. Evaluations were conducted in the Coin Game environment over $\kappa = 50$ steps, with a population size of $\eta = 250$ and $\mu = 200$ generations.

\begin{figure}[ht]
    \centering
    \includegraphics[width=\linewidth]{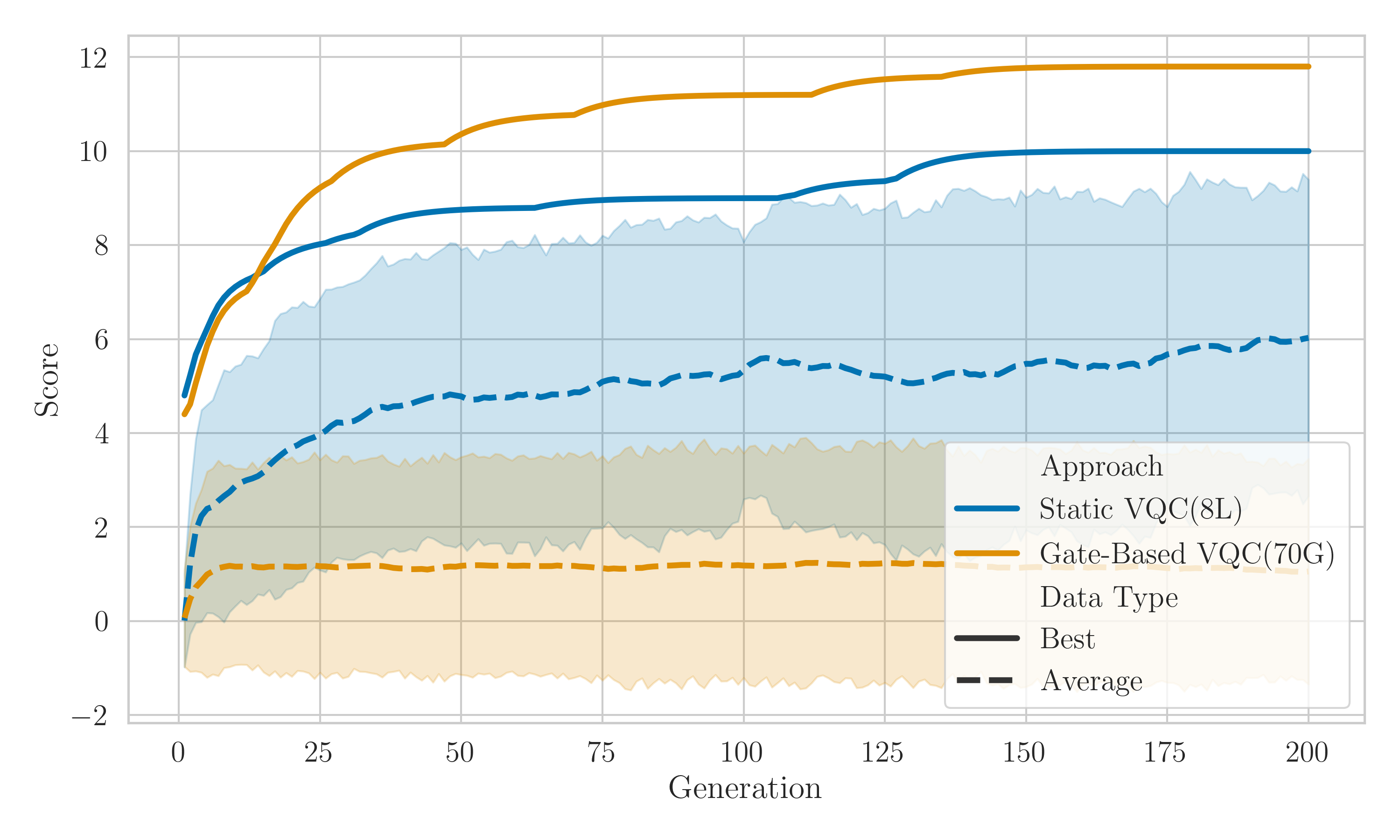}
    \caption{Best and average score of Gate-Based approach (with 70 initial gates) and static approach (without architectural changes) averaged over 5 seeds.}
    \label{fig:70VQC_Baseline}
\end{figure}

We start with the comparison of a 70 gate Gate-Based VQC and the static baseline. \cref{fig:70VQC_Baseline} shows the score progression over 200 generations. Despite the Gate-Based approach having a lower average score throughout, it yielded a better score for the best agent. The static baseline's best score starts at 4.8, rises sharply to 8.8 by generation 50, and reaches 10 by generation 140, plateauing afterward. The average score rises from 0 to 5 in the first 75 generations, slowly increasing to 6 by generation 200. The Gate-Based best score starts at 4.4, quickly reaches 10 by generation 40, and steadily grows to 11.8 by generation 150, remaining steady afterward. The average score rises to slightly above 1 by generation 6, staying constant until generation 200.

The Gate-Based best agent reached the static baseline’s top score of 10 a full 100 generations earlier and achieved a best agent score nearly 2 points higher within 200 generations. This suggests that evolving the architecture allows for exploring more optimal structures, leading to a more optimized solution candidate in fewer generations.

\begin{figure}[ht]
    \centering
    \includegraphics[width=\linewidth]{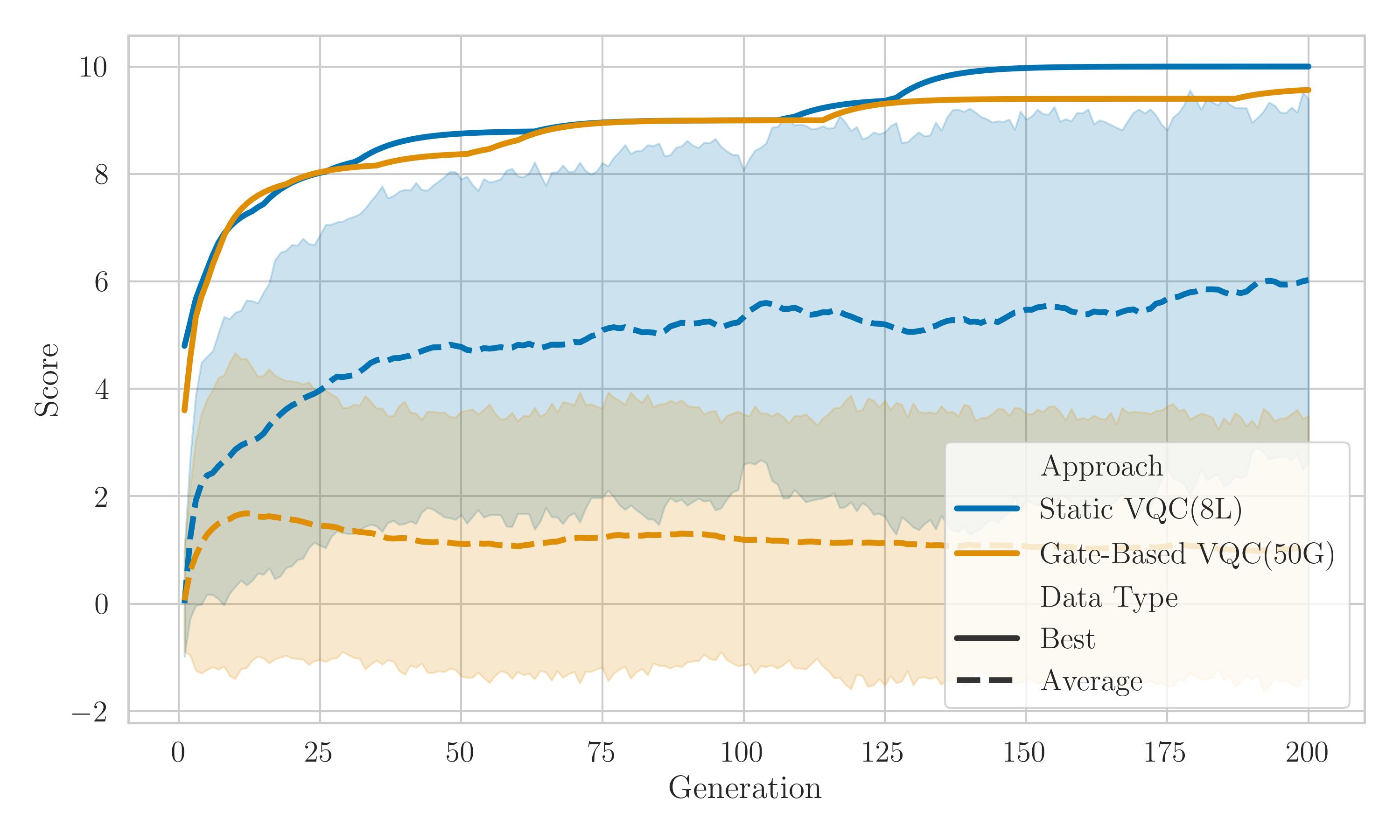}
    \caption{Best and average score of Gate-Based approach (with 50 initial gates) and static approach (without architectural changes) averaged over 5 seeds.}
    \label{fig:50VQC_Baseline}
\end{figure}

Now, we continue with the comparison of the 50 gate Gate-Based approach and the static baseline, where the results are depicted in \cref{fig:50VQC_Baseline}. We can identify the Gate-Based approach yielding comparable best scores to the static baseline. The static baseline, which consists of 192 gates distributed among 8 layers on 6 qubits, reaches a best score of 10, whereas the average score peaks at 6 by generation 200. The Gate-Based approach starts with 50 gates that are reduced to approximately 25 gates for the best agent by generation 120, due to the architectural changes implemented through evolutionary processes. The best agent scores similarly to the static baseline with a to score of 9.5. The average score peaks at 1.7 by generation 12 and then stabilizes around 1. 

Using fewer gates reduces computational resource requirements, which is crucial in the NISQ era. The Gate-Based approach’s computation time averaged 8.06 hours per seed compared to 27.13 hours for the fixed architecture, representing speedup of 70.29 \%.

The Gate-Based VQC with a flexible architecture achieves scores for the best candidate that are similar to those of the static baseline while operating more efficiently by using fewer gates and reducing computation time. This underscores the benefits of adopting a flexible architectural approach in VQCs, which not only improves performance but also reduces resource usage.
\section{Conclusion}
\label{sec: conclusion}
At the moment, MAQRL suffers from barren-plateaus and vanishing gradients due to their gradient based training \cite{chen2022variational}, \cite{franz2023uncovering}. K{\"o}lle et al. \cite{kolle2023multi} extended an evolutionary optimization process by Chen et al. \cite{chen2022variational} to a Multi-Agent setting and used different evolutionary strategies. They proposed three approaches: layer-wise crossover with mutation, random crossover with mutation and mutation only. The evaluation in a 3x3 grid world Coin Game showed, that their VQC approach performed significantly better than classic neural networks with a similar amount of trainable parameters. Comparing the VQC with a much bigger neural network resulted in a similar outcome. Here, it was possible to reduce the number of parameters by 97.88\%, still preserving a similar performance, which shows the effectiveness of VQC in MAQRL environments. 

We extended this approach using the Coin Game by proposing three different evolutionary strategies, namely a Gate-Based, a Layer-Based and a Prototype-Based approach. After testing recombination and mutation vs mutation only-strategies, which resulted in a significant better performance of mutation-only strategies, we continued with mutation-only in the next experiments. When comparing Gate-Based, Layer-Based and Prototype-Based approaches, we found that on average, all three strategies lead to a similar score, but focusing on the best agents reveals a better performance of the Gate-Based approach. The Gate-Based approach also achieves better values for the total coins collected, own coins collected and, own coin rate. Additionally, this approach needed fewest parameterized gates. When comparing Gate-Based VQCs of different gate sizes and a static baseline, the baseline meets higher score values on average, but the best baseline agents' score is almost equal to (50 gates) or even succumbs (70 gates) the best Gate-Based agent score. In summary we showed that we can save computational resources and still achieve good results at the same time using architectural evolutionary strategies.

For future work, the experiments could be run on real quantum hardware to determine if there are differences in the results. Another interesting application of the recombination strategies can be Large Language Models. 

\section*{Acknowledgements}
This research is part of the Munich Quantum Valley, which is supported by the Bavarian state government with funds from the Hightech Agenda Bayern Plus.

%\clearpage
\bibliographystyle{splncs04}
\bibliography{main}
\end{document}